\documentclass[aps,prx,floats,twocolumn,superscriptaddress]{revtex4-1}
\usepackage{amsmath}
\usepackage{amsfonts}
\usepackage{amssymb}
\usepackage{color}
\usepackage{enumerate}
\DeclareMathOperator{\sign}{sign}
\usepackage{ dsfont }
\newcommand{\beq}{\begin{equation}}
\newcommand{\eeq}{\end{equation}}
\newcommand{\beqa}{\begin{eqnarray}}
\newcommand{\eeqa}{\end{eqnarray}}
\newcommand{\pr}{^{\prime}}
\usepackage{graphicx,amssymb,amsmath,subfigure}
\usepackage{bm}

\newcommand{\braket}[1]{\left<#1\right>}

\begin{document}
\title{Parity Anomaly in the non-linear response of Nodal-Line Semimetals}
\author{Alberto Mart\'in-Ruiz}
\email{alberto.martin@c3.unam.mx}
\affiliation{Materials Science Factory, Instituto de Ciencia de Materiales de Madrid, CSIC, Cantoblanco; 28049 Madrid, Spain.}
\affiliation{Centro de Ciencias de la Complejidad, Universidad Nacional Aut\'{o}noma de M\'{e}xico, 04510 Ciudad de M\'{e}xico, M\'{e}xico}
\author{Alberto Cortijo}
\email{alberto.cortijo@csic.es}
\affiliation{Materials Science Factory, Instituto de Ciencia de Materiales de Madrid, CSIC, Cantoblanco; 28049 Madrid, Spain.}

\begin{abstract}
Nodal-line semimetals are topological semimetals characterized by one-dimensional band-touching loops protected
by the combined symmetry of inversion $\mathcal{P}$ and time-reversal $\mathcal{T}$ in absence of spin-orbit coupling. These nodal loops can be understood as
a one-parameter family of Dirac points exhibiting the parity anomaly associated to $\mathcal{P}*\mathcal{T}$ symmetry.
We find that the parity anomaly also appears in the non-linear optical response of these systems in an analogous way to the linear response transport. We analyze the presence of a tilting term in the Hamiltonian as an element that does not spoil $\mathcal{P}*\mathcal{T}$ symmetry: while it is $\mathcal{P}*\mathcal{T}$-symmetric, it breaks separately both $\mathcal{P}$ and $\mathcal{T}$ symmetries, allowing for the potential experimental observability of the linear and non-linear Hall conductivities in appropriate nodal-line semimetals. We also propose effective actions for both the linear and non-linear electromagnetic responses of tilted nodal-line semimetals. We find that the linear Hall-like response of tilted Nodal-line semimetals is an axion response due to the parity anomaly, extending the class of systems that display such electromagnetic response.
\end{abstract}
\maketitle
\section{Introduction}
\label{sec:intro}
Nodal line semimetals (NLSM) are three-dimensional semimetallic systems where the valence and conduction bands closest to the Fermi level cross each other at a one-dimensional lines in momentum space, in contrast to a discrete set of points, as it happens in Weyl semimetals\cite{BHB11,AMV18}. These semimetals are endowed with a topological $\mathbb{Z}_2$ invariant that provides topological stability of this band structure. In practical grounds, this $\mathbb{Z}_2$ topological charge is the Berry phase $\pi$ acquired by any curve in momentum space that is threaded by the nodal loop (if the curve is not threaded by the nodal loop, the Berry phase is simply zero). In absence of spin-orbit coupling (SOC), it is known that the combined symmetry of parity and time reversal symmetry $\mathcal{P}*\mathcal{T}$ is the symmetry that allows this topological protection\cite{FCK15}. In presence of SOC, the situation is richer and several combination of symmetries can stabilize the nodal loops\cite{GSM15,FCK15}.

Until now, the only experimental signature of the $\mathbb{Z}_2$ topological invariant appears in the detection of the Berry phase in magnetotransport measurements\cite{RK15,HTL16,ASG16,LB17,ESD17,ZGZ17,PDS17,SSJ17,MCK17,HML18,LWW18,ODC18}. Then it is relevant to explore other physical phenomena where this invariant can be observed. In this regard, it has been proposed that the $\mathbb{Z}_2$ topological charge could also leave its imprint in other transport coefficients\cite{RH17,B18,RZS18}. Indeed, the $\mathbb{Z}_2$ topological invariant can be understood as a three dimensional variant of the parity anomaly in two spatial dimensions\cite{B18,RZS18}. In the Quantum Field Theory context, the parity anomaly appears in two dimensional massless Dirac systems as the failure of keeping the $\mathcal{P}*\mathcal{T}$ symmetry after regularizing the theory in a gauge invariant way\cite{R84,R842}. In the effective electromagnetic action, it manisfests as a radiatively induced Chern-Simons term that gives rise to a finite, quantized anomalous Hall conductivity. Then, by noticing that both nodal loops in three dimensional $\mathcal{P}*\mathcal{T}$ symmetric NLSM and the nodal points in massless  $\mathcal{P}*\mathcal{T}$ symmetric Dirac fermions in two dimensions have codimension $d_c=1$, it is expected that they will share the same topological invariant\cite{FCK15,RZS18}. There is another equivalent, but more operational definition of anomaly: Given a theory displaying some symmetry, add to it some term that explicitly breaks this symmetry and compute \emph{sensible} quantities beyond the tree level (including quantum corrections). When sending back this term to zero, if these responses do not go to the responses obtained in absence of the symmetry breaking parameter, we say that there is a quantum anomaly. In this operational definition the enphasis is put on the adding-removing the symmetry breaking term and the need for regulators is hidden in the word \emph{sensible}: finite and gauge invariant. This is the definition of anomaly that we will use in the present work.

It has been stated in Refs.\cite{B18,RZS18} that NLSM exhibit a non zero Hall conductivity for each value of the polar angle $\phi$ that defines the one-dimensional loop in momentum space, so when considering all the values of $\phi$, the total Hall current vanishes:
\beq
\bm{J}_{H}=\frac{e^2}{2\pi^2}\sign(m)k_0\bm{E}\times\int d\phi \hat{\bm{e}}_{\phi}=0,\label{ZeroHall}
\eeq
where the vector $\hat{\bm{e}}_{\phi}=-\sin\phi \hat{\bm{x}}+\cos\phi \hat{\bm{y}}$ is the unitary polar vector in cylindrical coordinates in momentum space. Even at finite $\mathcal{P}*\mathcal{T}$ symmetry breaking mass term, the angular integration of (\ref{ZeroHall}) gives zero. The reason behind the cancellation of the Hall conductivity in Eq.(\ref{ZeroHall}) is because each point at the nodal line has another point in the nodal loop related by inversion symmetry, so, when summing over all the possible points labeled by $\phi$, each pair of points related by inversion symmetry will contribute oppositely to the Hall conductivity, explaining the result (\ref{ZeroHall})\cite{NN81}. The scenario is reminiscent to what happens in graphene, where there are two nodal points in the Brillouin Zone related by inversion symmetry. Adding the same symmetry breaking parameter in both nodes breaks inversion but not time reversal symmetry, so the system displays a valley Hall effect (or spin Hall effect when considering spin) instead of an anomalous Hall effect\cite{KM05}. If a symmetry breaking mass with opposite sign is added to each node, then a quantum anomalous Hall effect is obtained\cite{H88}.
Nevertheless, for any fixed angle $\phi$, the corresponding point in the nodal loop displays the parity anomaly, and using the standard bulk-boundary correspondence associated to the Chern-Simons term in two spatial dimensions, we can use the parity anomaly for each value of $\phi$ to account for the drumhead surface states appearing in NLSM\cite{CCC16,MG18}.
The purpose of the present work is twofold: first, we study the non-linear Hall conductivity in NLSM to see how the parity anomaly also extends to non-linear effects, along the lines of previous studies\cite{DGI09,SF15}. Second, we analyze the way to add terms that, respecting the $\mathcal{P}*\mathcal{T}$ symmetry, might alter the balance between inversion-related points in the nodal loop, and obtain non-vanishing results for the linear Hall conductivity (\ref{ZeroHall}) and for the non-linear Hall responses.

In Section \ref{sec:model} we describe the effective $\bm{k}\cdot\bm{p}$ two-band model for NLSM including a tilting term, describing also its properties under $\mathcal{P}$ and $\mathcal{T}$ symmetries. In Sec. \ref{sec:linear} we review the fate of the linear Hall conductivity for NLSM in absence of the tilting term, and how this precise term allows for a non-vanishing Hall conductivity, implying that in principle the term coming from the parity anomaly could be observed. In Sec.\ref{sec:nonlinear} we study the Berry curvature dipole moment which is the element that governs the Hall-like non-linear optical response in absence of interband transitions. In Sec.\ref{sec:interband} we study the non-linear response associated to interband transitions and their relation with the parity anomaly. We end by commenting the obtained results in Sec. \ref{sec:conclussions}.

\section{Model and preliminaries} 
\label{sec:model}
In the present work we will use the following simplified $\mathcal{P}*\mathcal{T}$ symmetry breaking model\cite{CCC16} (throughout this work we will use $\hbar=1$):
\beq
H(\bm{k})= \sigma_0\bm{v}_{t}\cdot\bm{k}+\frac{1}{\Lambda}(k^2_0-k^2_{\rho})\sigma_1+ m\sigma_2+ v_3 k_3 \sigma_3,\label{Ham1}
\eeq
where $k^2_{\rho}=k^2_1+k^2_2$. The Pauli matrices $\bm{\sigma}$ represent an effective orbital basis, not necessarily the spin degree of freedom\cite{GSM15}. The parameter $\Lambda$ is a momentum scale that comes from any particular lattice realization of the $\bm{k}\cdot\bm{p}$ model (\ref{Ham1}). Alternatively, one can work with a four-band model that is linear in all the momenta (this situation corresponds to systems with SOC), but the relevant bands will be these that are crossed by the Fermi level, leaving with an anisotropic two-band model and a parameter $\Lambda$ depending on the SOC coupling constant (among other lattice parameters), after projecting out the high energy sector. Since the physics described in this work does not depend on this parameter, in the rest of the paper, we will make non dimensional variables absorbing this extra model-dependent parameter. The presence of the mass term $m\sigma_2$ breaks $\mathcal{P}*\mathcal{T}$ symmetry. When $m$ and $\bm{v}_{t}$ are zero, the model consists in two bands that touch each other at a one-dimensional ring in momentum space (located in the $k_3=0$ plane). In this case ($m=|\bm{v}_{t}|\equiv v_{t}=0$) the system is not only invariant under the combined $\mathcal{P}*\mathcal{T}$ symmetry, it is also symmetric under both $\mathcal{P}$ and $\mathcal{T}$ symmetries separately. However, the topological stability of the nodal line comes from the invariance of this combined symmetry, so any term in the Hamiltonian $H(\bm{k})$  invariant under $\mathcal{P}*\mathcal{T}$ will not alter the topological protection of $H(\bm{k})$. This is precisely the case of the term $\sigma_0\bm{v}_{t}\cdot\bm{k}$. This shift in energies has been used in graphene to study the half-integer contribution of each Dirac point to the quantum Hall effect\cite{WHA10}, and tilted nodal loops have been predicted to occur in alkaline-earth stannides, germanides, and silicides\cite{HLV16} and in other materials displaying non-symmorfic symmetries\cite{HJL17,AWJ17}. In the orbital basis employed in the model (\ref{Ham1}), the inversion and time reversal symmetries $\mathcal{P}$ and $\mathcal{T}$ are implemented by the operators $P=\sigma_3$ and and $T=\mathcal{K}\sigma_3$ respectively ($\mathcal{K}$ stands for complex conjugation and for both symmetries the momentum $\bm{k}$ has to reverse its direction), so $\mathcal{P}*\mathcal{T}$ symmetry is implemented by $PT=\mathcal{K}$, that is, the reality condition for the Hamiltonian $H(\bm{k})$: $H^{*}(\bm{k})=H(\bm{k})$. From this it is clear that the term $\sigma_0\bm{v}_{t}\cdot\bm{k}$ breaks time reversal and inversion symmetries separately (due to the flip of the sign of $\bm{k}$) but leaves invariant $\mathcal{P}*\mathcal{T}$ symmetry. As mentioned in the introduction, the topological invariant associated to the loop remains intact despite the presence of $\sigma_0\bm{v}_{t}\cdot\bm{k}$. 
\begin{figure}
\includegraphics[scale=0.2]{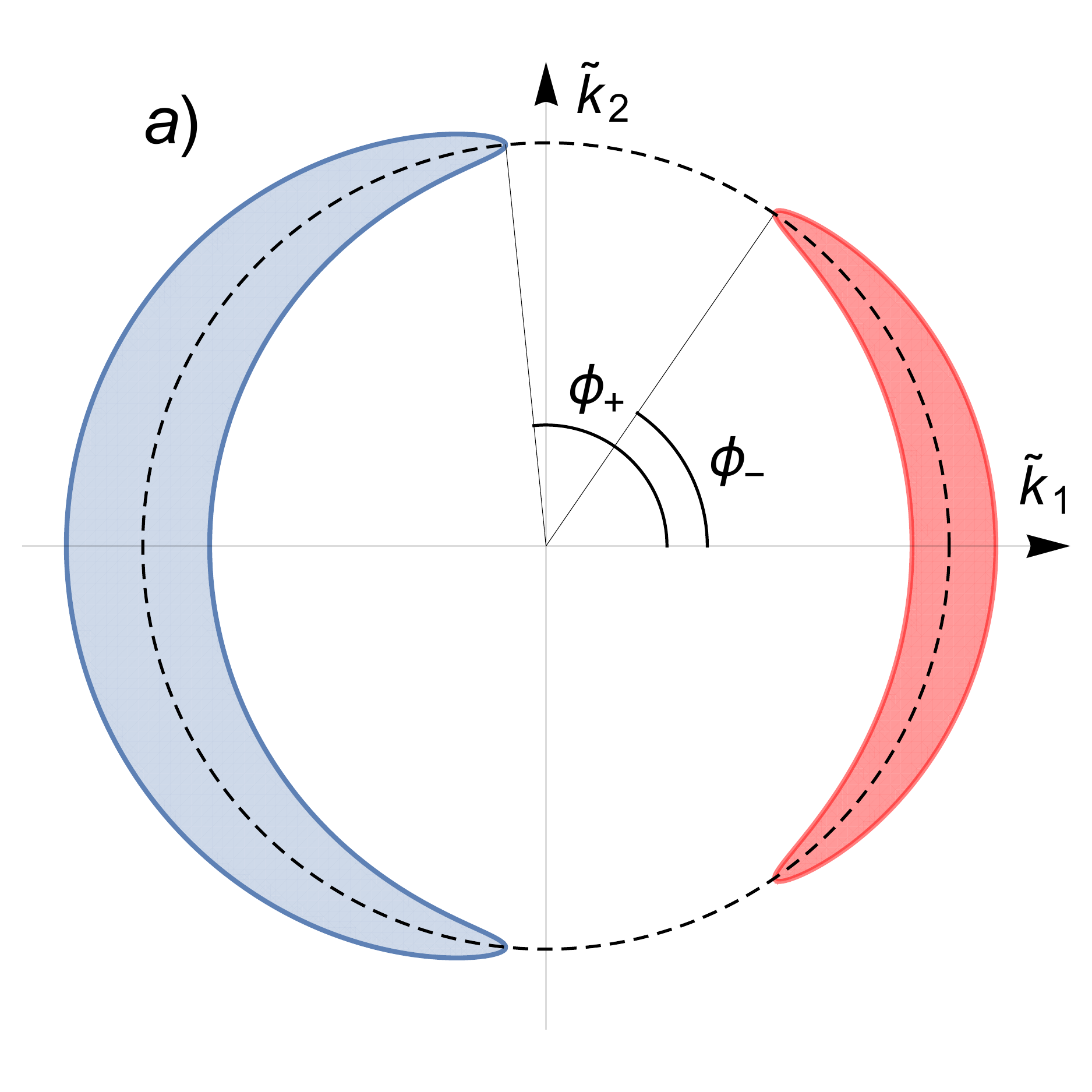}
\includegraphics[scale=0.2]{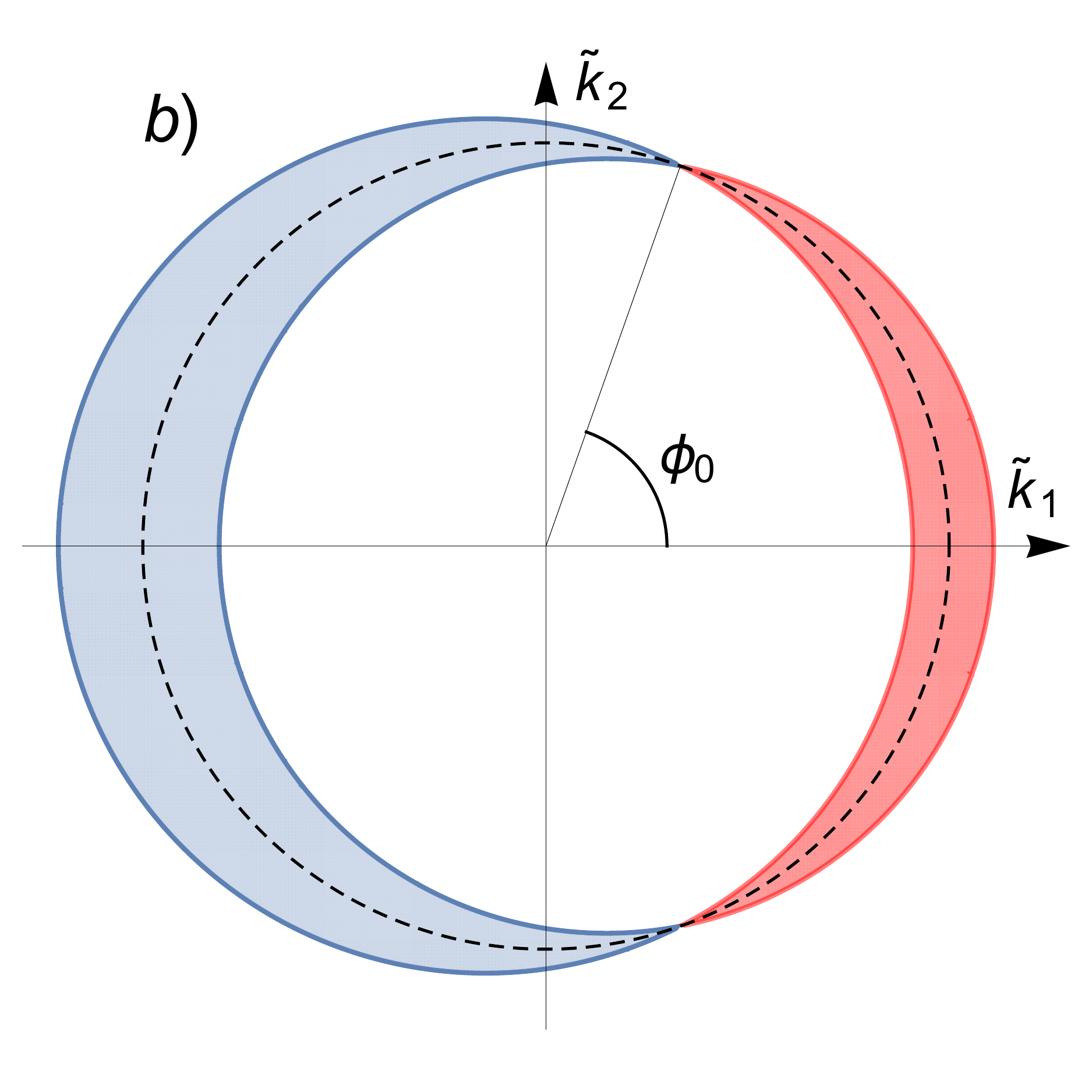}
\caption{(color online) Cross sections of the Fermi surface for $k _{3} = 0$, for fixed chemical potential and tilt ($\mu < v _{t}k_0$) for the case of $\bm{v}_t=v_t \hat{\bm{e}}_1$ ($\alpha=0$). We observe that the separation between the conduction and valence band closes as $m \rightarrow 0$, as expected. Panel (a) shows the generic case for $m\neq0$. At $m=0$ (panel (b)) the two bands touch each other at points over the loop, defining the integration limits of the integral of the polar angle in momentum space.}
\label{cross}
\end{figure}

\section{Linear Hall current} 
\label{sec:linear}
\begin{figure*}
\includegraphics[scale=0.35]{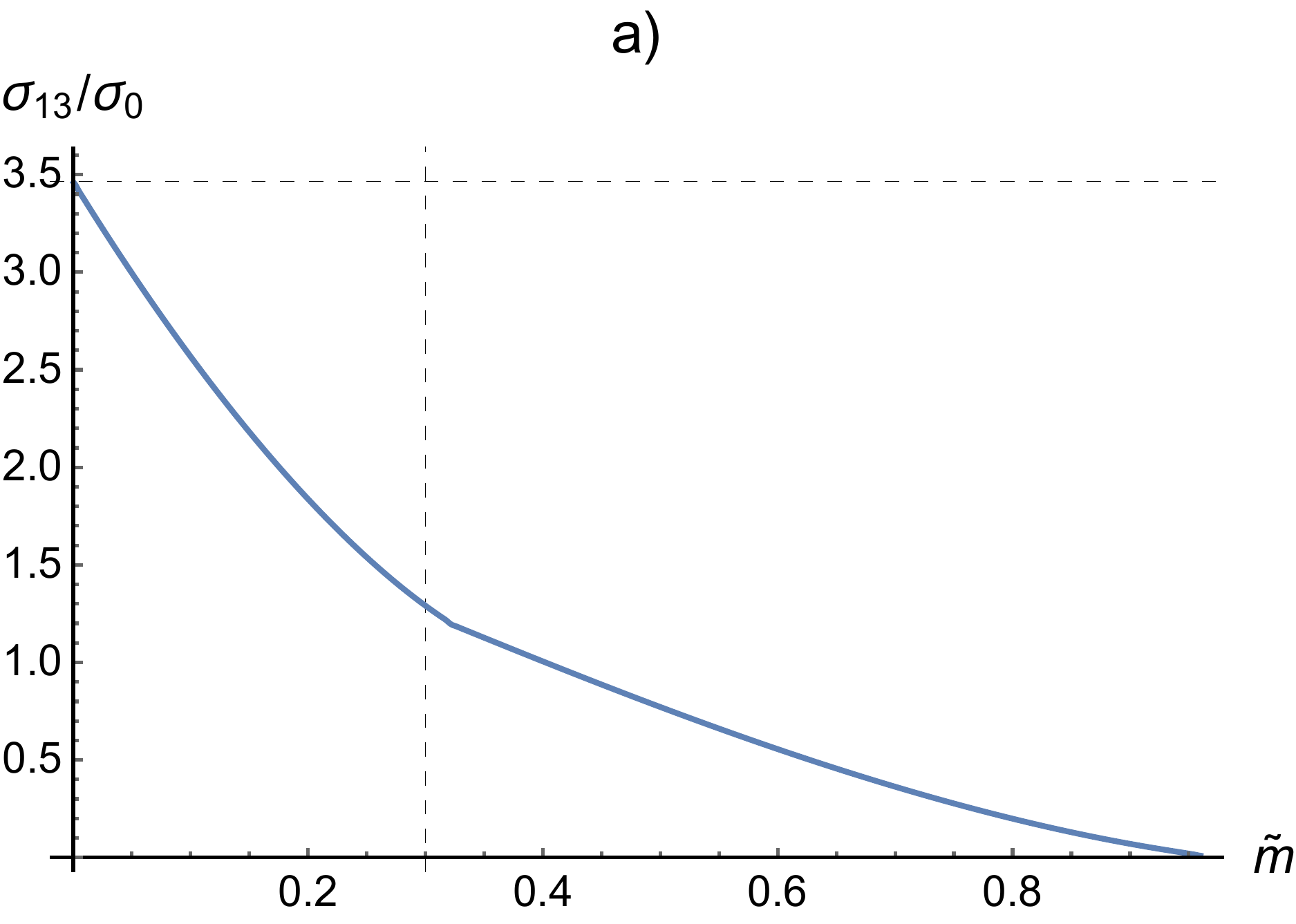}
\includegraphics[scale=0.35]{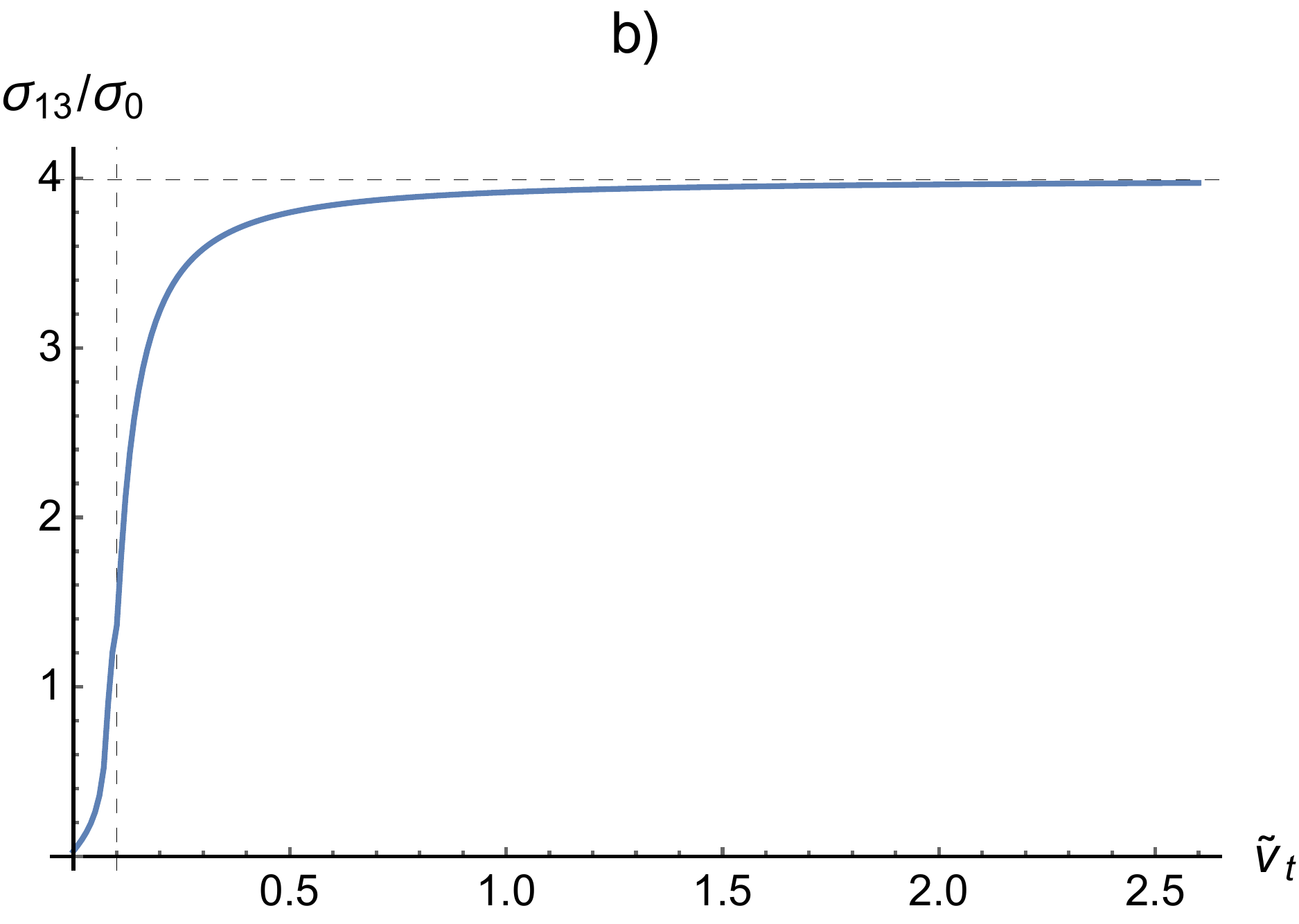}
\caption{Evolution of the conductivity $\sigma _{13}$ (in units of the quantum of conductance $\sigma _{0}=e^2/h$) a) as a function of the dimensionless mass gap $\tilde{m}\equiv m\Lambda /k^2_0$ for $\tilde{\mu}\equiv \mu\Lambda/k^2_0= 0.3$ and $\tilde{v} _{t} \equiv v_t\Lambda /k_0= 0.6$ and b) as a function of the tilt $\tilde{v} _t $ for fixed $\tilde{m} = 0.01$ and $\tilde{\mu} = 0.1$.}
\label{HallCurrentNumerical}
\end{figure*}
Let us analyze how the presence of the term $\sigma_0\bm{v}_{t}\cdot\bm{k}$ modifies the Hall conductivity in Eq.(\ref{ZeroHall}). This analysis is not useful only to prepare us for the non-linear optical/transport responses, but it is interesting by its own, since the orientation of the loop in momentum space will play a crucial role for the detection of the Hall effect in NLSM, as we will see. 

The part of the electric current density (as a response to an electric field $\bm{E}$) associated to the parity anomaly can be computed, for a two-band model, with the following expression:
\beq
\bm{J}_{H}=\sum_{s=\pm}\frac{e^2}{8\pi^3}\int d^3\bm{k}f^{s}_0(\bm{k})\bm{E}\times\bm{\Omega}^s_{\bm{k}},\label{Halldef}
\eeq
where $\bm{\Omega}^s_{\bm{k}}$ and $f^{s}_0(\bm{k})$ are the Berry curvature and the equilibrium distribution function associated to the band $s$, respectively. We have computed the current (\ref{Halldef}) for zero and non-zero mass term $m$. The Berry curvature for the Hamiltonian (\ref{Ham1}) has been computed in Ref.\cite{RZS18}. Let us focus first on the $m=0$ limit that is amenable of analytical tractability. In this limit, the Berry curvature of the band $s$ can be written as 
\beq
\bm{\Omega}^s_{\bm{k}}=s\pi\sign(m)\delta(k_3)\delta(k_{\rho}-k_0)\hat{\bm{e}}_{\phi},\label{Berryzero}
\eeq
being independent of the tilt parameter $\bm{v}_t$. Due to the presence of the tilting vector $\bm{v}_{t}$, it is expected that for some range of chemical potentials, both conduction and valence bands will contribute to (\ref{Halldef})\cite{AMM17}. In Fig. \ref{cross}(a) we plot a cross section of the Fermi surface for $k _{3} = 0$ and a given nonzero value of the symmetry breaking parameter $m$. When $m=0$, as shown Fig. \ref{cross}(b), the electron and hole Fermi surfaces touch each other at two points over the nodal loop, defining the integration limits of the integral of the polar angle in momentum space.

For $k_3=0$ and $k_{\rho}=k_0$, the zero-temperature equilibrium distribution functions for electrons and holes only depend on the polar angle in momentum space: $f^{s}_0=\Theta[s(\mu-v_t k_0\cos(\phi-\alpha))]$, where $\alpha$ is the angle between the vector $\bm{v}_t$ and the horizontal axis. From these expressions, it is easy to see that the integration over the polar angle will not run over all the angles defining the nodal loop, but they will depend on the ratio $\mu/v_t k_0$ and the angle $\alpha$, as shown in Fig.\ref{cross}.  

Performing the angular integration, the Hall current takes the form (the details are presented in the Appendix \ref{appendix:Hallcurrent}):
\beq
\bm{J}_{H}=\frac{e^2}{2\pi^2}\sign(m)k_0\sqrt{1-\frac{\mu^2}{v^2_t k^2_0}}\bm{E}\times(\hat{\bm{v}}_t\times\hat{\bm{t}}),\label{Halltilted}
\eeq
where $\hat{\bm{t}}$ is the unit vector perpendicular to the plane containing the nodal loop in momentum space, and $\hat{\bm{v}}_t=\bm{v}_t/v_t$.
We see that, while the chemical potential does not exceed $v_t k_0$ (physically, the conduction band does not extend over the whole nodal loop), we obtain a non-vanishing Hall current. While the Berry curvature still corresponds to the $\mathcal{P}*\mathcal{T}$ situation, the effect of the time reversal symmetry breaking tilting term  (still preserving $\mathcal{P}*\mathcal{T}$) is to reduce the integration region over the Hall conductivity is constructed.

The current (\ref{Halltilted}) is obtained by using the $m\to 0$ limit of the Berry curvature $\bm{\Omega}^s_{\bm{k}}$. We have also computed the $m\neq 0$ situation numerically to observe the evolution of the current $\bm{J}_{H}$ for small yet finite values of $m$. The relation of the current $\bm{J}_H$, the electric field, and the vectors $\hat{\bm{v}}_t$ and $\hat{\bm{t}}$ is still valid as written in Eq.(\ref{Halltilted}). In Fig. \ref{HallCurrentNumerical}(a) (Fig. \ref{HallCurrentNumerical}(b)) we plot  the component $\sigma_{13}$ as a function of $m$ at fixed $\mu$ and tilt $v_{t}$  (as a function of tilt $v_t$ at fixed $\mu$ and $m$). 

It is interesting to observe that there is a value of the mass $m$ above it the conductivity goes to zero, even for a non-zero tilt. We can understand this behavior by considering the Fermi distributions (\ref{distelectrons}) and (\ref{distholes}) at non zero values of $m$. As it can be seen in Fig. \ref{cross}, the integration contour in (\ref{Halldef}) is defined by the values of the Fermi distribution functions for valence and conduction bands over the nodal loop. For the angle $\phi$, we have the condition $-1\leq \cos\phi_s\leq 1$ with $s=1$ ($s=-1$) for electrons (holes). From the dispersion relations (\ref{tiltdispersions}) over the nodal loop, this condition translates into the inequalities (again, for $s=\pm1$):
\beq
-(v_t k_0-s\mu)\leq m\leq(v_t k_0+s\mu).\label{inequalities}
\eeq
We can assume that $v_t k_0>\mu$ (otherwise the conduction band covers all the angles $\phi$ and the valence band does not contribute, obtaining a vanishing result as it happens in (\ref{ZeroHall})). It is easy to see that, if we want both bands to contribute to the integration in (\ref{Halldef}), both inequalities (\ref{inequalities}) must simultaneously hold, restricting the possible values of $m$ to the interval $[0,v_t k_0+\mu)$, explaining the result plotted in Fig. \ref{HallCurrentNumerical}(a).

We finish this section discussing which type of electromagnetic action gives rise to the current (\ref{Halltilted}).  Since we are working in the infinite space case (no boundaries) the Hall current (\ref{Halltilted}) can be written as the functional derivative of an effective action term $\Gamma^{(1)}_{H}$, $J_{H,a}=\delta \Gamma^{(1)}_{H}/\delta A_{a}$ (we write $C^*=\frac{e^2}{2\pi^2}\sign(m)k_0\sqrt{1-\frac{\mu^2}{v^2_t k^2_0}}$ for simplicity) that takes the gauge invariant form:
\beq
\Gamma^{(1)}_{H}=\int d^3 \bm{x}dt C^* \epsilon^{\mu\nu\rho\sigma}q_{\mu}A_{\nu}\partial_{\rho}A_{\sigma},\label{effaction1}
\eeq
where we have defined the vector $q_{\mu}=(0, k_0\hat{\bm{v}}_t\times\hat{\bm{t}})$, having dimensions of inverse length. The main difference with the proposals available in the literature for untilted NLSM is that there is an explicit vector $q_{\mu}$ that does not vanish at the end of the calculation\cite{B18,RZS18}.

We identify this action as the axion electromagnetic action in three spatial dimensions, and conclude that, tilted NLSM display an axion term in their electromagnetic response due to the parity anomaly, in contrast to Weyl semimetals, where the the axion term appears due to the chiral anomaly\cite{ZB12,CWB13}. Also, the situation is different to the case of axion insulators: Here we are dealing with a gapless system displaying the parity anomaly\cite{QHZ08}.

\section{Non-linear Hall current}
\label{sec:nonlinear}

In systems lacking inversion symmetry, there is a non-linear analog to the previously discussed anomalous Hall current\cite{MO10,DGI09,SF15,ZSY18}. In three dimensional Weyl and Dirac semimetals, the generated photocurrents have attracted attention due to their connection with the chiral anomaly\cite{C16,MZO16,ZSY18,RP18}. Under the effect of an oscillating electric field, $\bm{E}=e^{i\omega t}\bm{\mathcal{E}} + e^{-i\omega t} \bm{\mathcal{E}} ^{\ast}$, two non-linear currents can be generated: a photogalvanic current $\bm{J} _{a} ^{0} = \boldsymbol{\mathcal{J}} _{a} ^{(0)} +\boldsymbol{\mathcal{J}} _{a} ^{(0) \ast}$ and a second harmonic current, $\bm{J} _{a} ^{2 \omega} = e ^{2i \omega t}  \boldsymbol{\mathcal{J}} _{a} ^{(2 \omega)} + e ^{ - 2i \omega t}  \boldsymbol{\mathcal{J}} _{a} ^{(2 \omega) \ast}$, where 
\begin{align}
\boldsymbol{\mathcal{J}} _{a} ^{(0)} =\chi_{abc}(\omega)\mathcal{E}_{b}\mathcal{E}^{*}_{c} , \quad \boldsymbol{\mathcal{J}} _{a} ^{(2 \omega )} =\chi_{abc}(\omega)\mathcal{E}_{b}\mathcal{E}_{c} e ^{2i \omega t} . \label{def}
\end{align}
Both currents are proportional to the same non-linear conductivity tensor $\chi^{s}_{abc}(\omega)$ that reads, in the collisionless limit ($\omega\tau\gg1$) as
\beq
\chi^{s}_{abc}(\omega)=i \epsilon_{adc}\frac{e^3}{\omega} \int\frac{d^3\bm{k}}{8\pi^3}f^s_{0}\partial_{b}\Omega^{s}_{d},\label{NLresp}
\eeq
from which it is customary to define the Berry curvature dipole moment (BCDM):
\beq
D_{ab}=\int\frac{d^3\bm{k}}{8\pi^3}f^s_{0}\partial_{a}\Omega^{s}_{b}.\label{dipole}
\eeq
In what follows we will devote ourselves to the analysis of $D_{ab}$, instead of the full expression (\ref{NLresp}). In the case of zero tilting, we will analyze the BCDM keeping finite the breaking parameter $m$ and performing the limit $m\to 0$ at the end of the calculations.
\begin{figure*}
\includegraphics[scale=0.3]{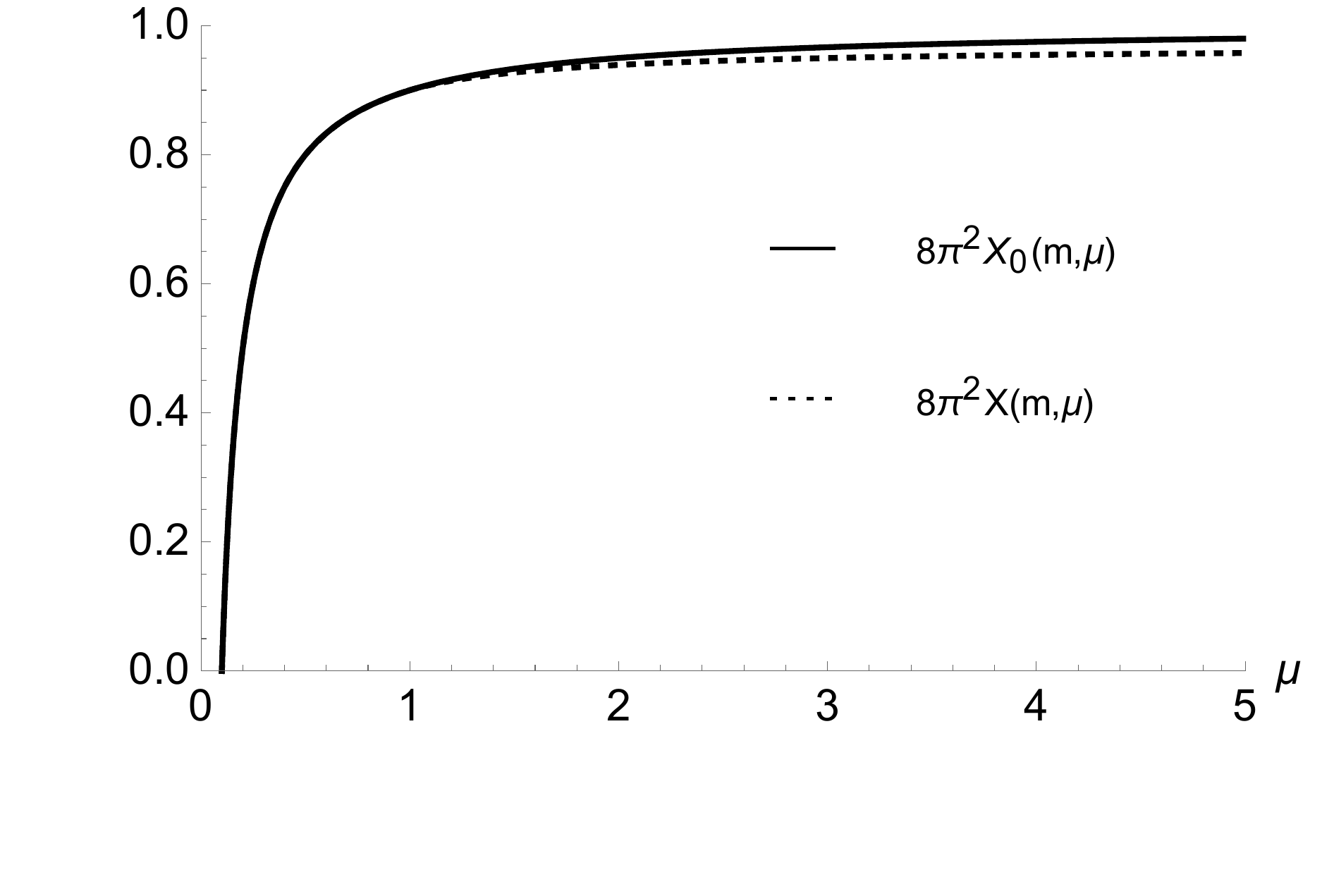}
\includegraphics[scale=0.3]{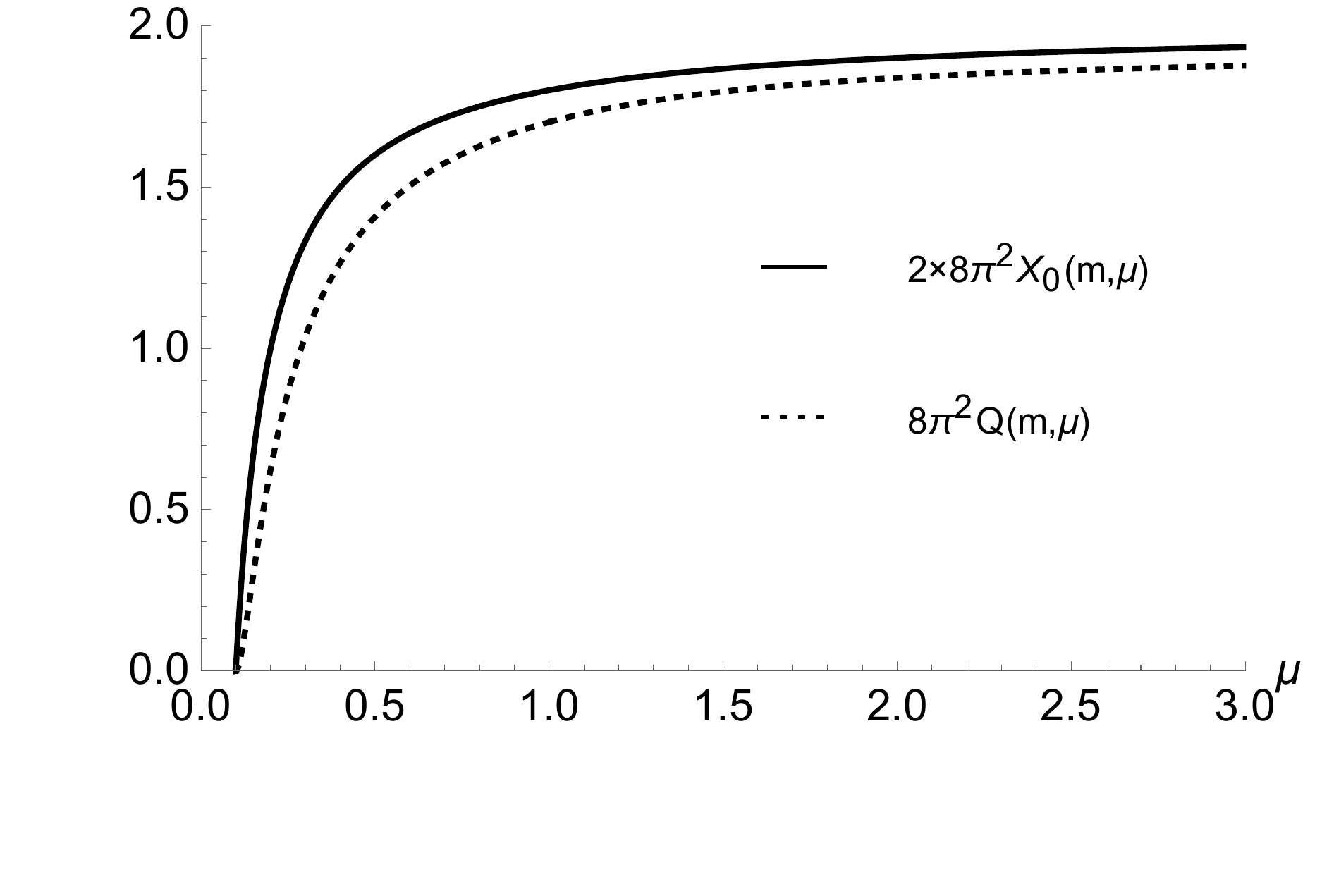}
\includegraphics[scale=0.3]{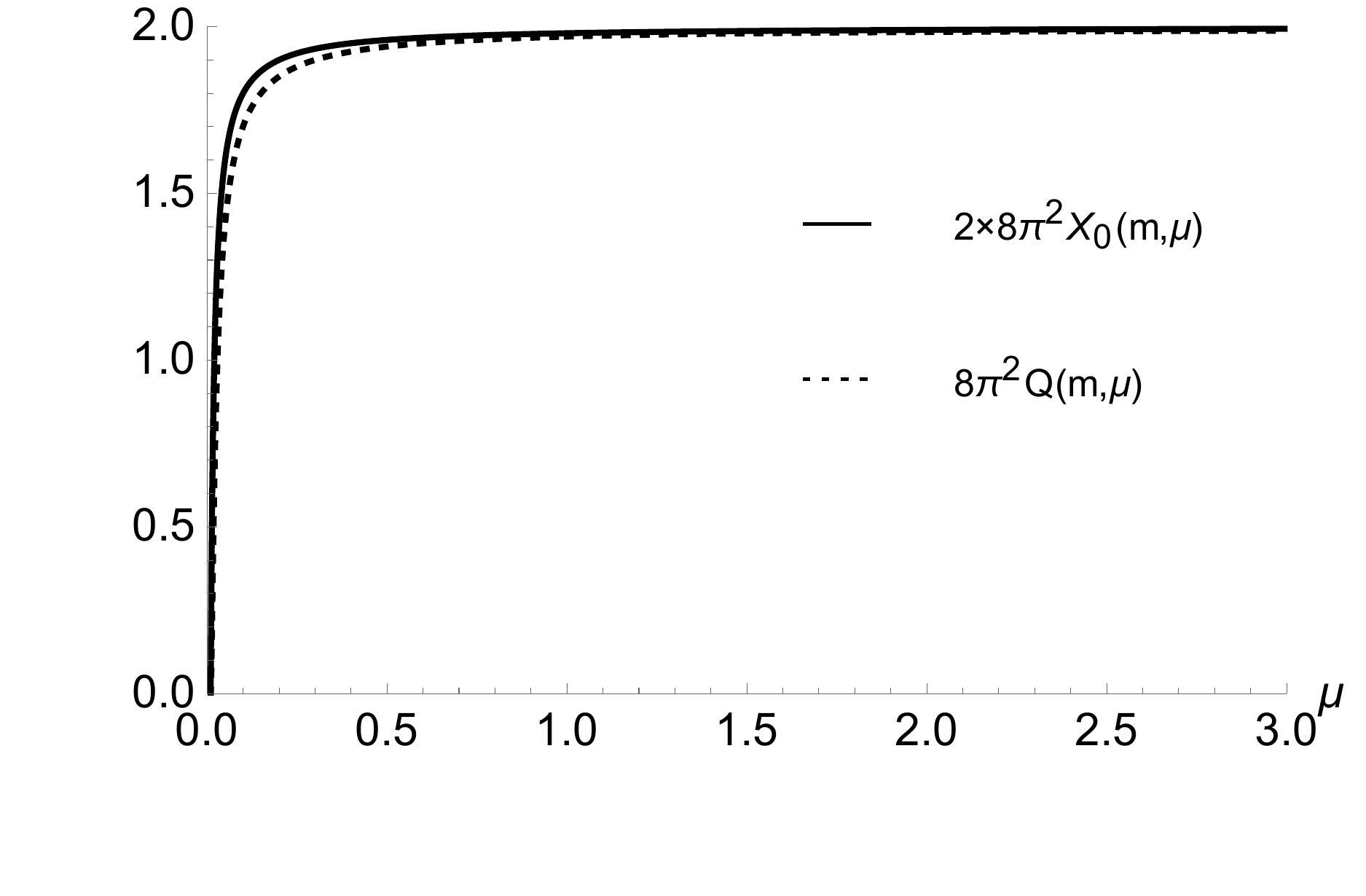}
\caption{Plot of the functions $8 \pi ^{2} X _{0} (m, \mu)$ (continuous line) and $8 \pi ^{2} X (m, \mu)$ (dashed line) for $\tilde{m} = 0.1$ (left). Plot of the functions $8 \pi ^{2} Q (m , \mu)$ (dashed line) and $2 \times 8 \pi ^{2} X _{0} (m , \mu)$ (continuous line) as a function of the dimensionless chemical potential $\tilde{\mu}$ for $\tilde{m} = 0.1$ (middle) and $\tilde{m} = 0.01$ (right).} \label{fig:PlotQfunc}
\end{figure*}
\subsection{Zero tilting}
For finite $m$ term, the Berry curvature is ($s=\pm 1$):
\beq
\bm{\Omega}^s_{\bm{k}}=s \frac{m v_3 k_\rho}{\Lambda \varepsilon ^3_{\bm{k}}}\hat{\bm{e}}_{\phi},\label{BerryC}
\eeq
where $\varepsilon_{\bm{k}}=\sqrt{m^2+v^2_3 k^2_3+(k^2_0-k^2_\rho)^2/\Lambda^2}$ is the dispersion relation for the conduction band (for the valence band, $\varepsilon^{-}_{\bm{k}}=-\varepsilon^{+}_{\bm{k}}=-\varepsilon_{\bm{k}}$) at zero tilting, and $k_{\rho}$ is again the radial momentum in cylindrical coordinates.

Since $\Omega_3=0$, we directly have $D_{a3}=0$. Also, by inspection, it is easy to see that $D_{3a}$ vanishes as well (the integrand is odd in $k_3$). By symmetry, we also have that $D_{11}=-D_{22}$. 

In parallel to the discussion of the Hall current done in Ref.\cite{RZS18}, it is convenient to define first the one-parameter family of BCDM parametrized by the polar angle, $D_{ab}=\int d\phi D^{\phi}_{ab}$, and write $D^{\phi}_{ab}$ in terms of two functions depending only on the mass term and the chemical potential:
\begin{subequations}
\beq
D^{\phi}_{11}=-D^{\phi}_{22}=\sin\phi\cos\phi Q(m,\mu),
\eeq
\beq
D^{\phi}_{12}=X(m, \mu)-\cos^2\phi Q(m,\mu),
\eeq
\beq
D^{\phi}_{21}=-X(m, \mu)+\sin^2\phi Q(m,\mu).
\eeq
\end{subequations}
Without loss of generality, in absence of the tilting term, we will assume that the chemical potential $\mu$ crosses the conduction band (meaning that we will use the band $s=1$), so the functions $Q$ and $X$ are defined as follows (the origin of these two functions can be seen in Appendix \ref{appendix:XandQ})
\beq
X(m,\mu)=\frac{m v_3}{8\pi^3\Lambda}\int dk_{\rho}dk_3 f_0(\varepsilon_{\bm{k}}) \frac{k_\rho}{\varepsilon^3_{\bm{k}}},\label{funcX}
\eeq
\beq
Q(m,\mu)=-\frac{6m v_3}{8\pi^3\Lambda^3}\int dk_{\rho}dk_3 f_0(\varepsilon_{\bm{k}}) \frac{k^3_\rho(k^2_0-k^2_{\rho})}{\varepsilon^5_{\bm{k}}}.\label{funcQ}
\eeq
The integration domain is determined by the Fermi-Dirac distribution function $f_0(\varepsilon_{\bm{k}})$, as usual. 

It is illuminating to consider the situation when the chemical potential lies slightly above the bandgap. In this case, only momenta close to the nodal loop will contribute to the integrals. In this limit, the dispersion can be approximated by $\varepsilon_{\bm{k}}\simeq\sqrt{m^2+v^2_3 k^2_3+(2k_0 q_\rho/\Lambda)^2}$, with $q_{\rho}=k_{\rho}-k_0$. Under this approximation, the function $X(m,\mu)$ can be computed analytically ($\mu\geq m$):
\beq
X(m,\mu)\to X_0(m,\mu)=\frac{1}{8\pi^2}\sign(m)\left(1-\frac{|m|}{\mu}\right).\label{funcX0}
\eeq

This expression is the same that appears in the linear Hall current (\ref{ZeroHall}) before taking the $m\to 0$ limit. In the zero mass limit, we obtain a finite value, which is a fingerprint of the parity anomaly. We have also evaluated numerically the full expression (\ref{funcX}) to check the validity of (\ref{funcX0}). In Fig.\ref{fig:PlotQfunc}(left) we have plotted both $X(m,\mu)$ and $X_0(m,\mu)$. For small values of the chemical potential and the mass term, both expressions are in agreement. Also, we can understand the behavior of $D_{12}$ as a function of $m$ plotted  Fig. \ref{fig:PlotQfunc}(right) along the lines discussed for the linear case at the end of Sec. \ref{sec:linear}.

Contrary to $X(m,\mu)$, the function $Q(m,\mu)$ is not peaked around the nodal loop so we have to evaluate it numerically. In Fig. \ref{fig:PlotQfunc}(middle, right) we have plotted $Q(m,\mu)$ for two values of the parameter $m$, and compared with $X_0(m,\mu)$. Crucially, we observe that when the mass parameter becomes smaller, the function $Q(m,\mu)$ approaches to $2X_0(m,\mu)$. This implies that the function $Q(m,\mu)$ also becomes non-zero in the zero-mass limit due to the parity anomaly.

We can write the $\phi$-dependent BCDM components in a compact form by using the unit vector normal to the nodal loop $\hat{\bm{t}}=(0,0,1)$ and the unitary vector perpendicular to $\hat{\bm{t}}$, $\hat{\bm{r}}=(\cos\phi, \sin\phi,0)$:
\beq
D^{\phi}_{ab}=\epsilon_{abc}\hat{t}_c X(m,\mu)+\epsilon_{bcd}\hat{r}_a\hat{r}_c \hat{t}_{d} Q(m,\mu).\label{phidipole}
\eeq

Now we see that the situation is similar but somewhat different to the linear case. As we mentioned, the expression in Eq.(\ref{ZeroHall}) is zero even for finite $m$ after integrating over $\phi$. In Eq.(\ref{phidipole}) we can integrate over $\phi$ at finite $m$ obtaining a \emph{non-zero result}: 
\beqa
D_{ab}(m,\mu)&=&\int D^{\phi}_{ab}d \phi \nonumber\\
&=&\pi \left[ 2X(m,\mu)-Q(m,\mu) \right] \epsilon_{abc}\hat{t}_c.\label{massdipole}
\eeqa
This is in agreement with the symmetry analysis performed in Ref.\cite{SF15}. Only the antisymmetric part of the matrix $D_{ab}$ survives to the integration, so in three spatial dimensions, this antisymmetric pseudotensor transforms as a polar vector, in our case the normal to the nodal loop $\hat{\bm{t}}$ playing such role. This also suggest the possibility that, for $m\neq 0$, the NLSM might become ferroelectric.

Let us consider now the $m=0$ case. We have seen that in this limit, the function $Q(m,\mu)$ approaches to $2X_0(m,\mu)$, taking the function $X_0(m,\mu)$ the exact value $X_0(0,\mu)=\frac{1}{8\pi^2}$. This means that each component of $D^{\phi}_{ab}$ is non-zero, but after integrating over $\phi$ we find a vanishing value for $D_{ab}(0,\mu)$ after using in (\ref{massdipole}) the fact that $Q(0,\mu)=2X(0,\mu)$.

\subsection{Non-zero tilting}

Let us compute the BCDM in presence of the tilting term. As we have seen in Sec. \ref{sec:linear}, this term does not alter the expression of the Berry curvature $\bm{\Omega}^s_{\bm{k}}$ but strongly modifies the integration contours for electrons and holes through the modification of the dispersion relations,
\beq
\varepsilon^s_{\bm{k}}=\bm{v}_t\cdot\bm{k}+s\sqrt{v^2_3 k^2_3+m^2+\frac{1}{\Lambda^2}(k^2_0-k^2_{\rho})^2}.\label{tiltdispersions}
\eeq
Previously, in absence of tilting, it was possible to write down simple expressions for the BCDM in terms of the functions $X(m,\mu)$ and $Q(m,\mu)$, since in that case only the conduction band contributed to the integral ($\mu>m$). We were then able to take the $m=0$ limit for these expressions and find an analytical expression that allowed us to compare with the parity anomaly in the linear Hall current.

A finite tilting makes both bands to contribute to the BCDM and the calculations at finite mass become more involved, precluding any analytical treatment. For this reason, we will analytically compute the BCDM at finite tilting directly at $m=0$. 
\begin{figure*}
\includegraphics[scale=0.3]{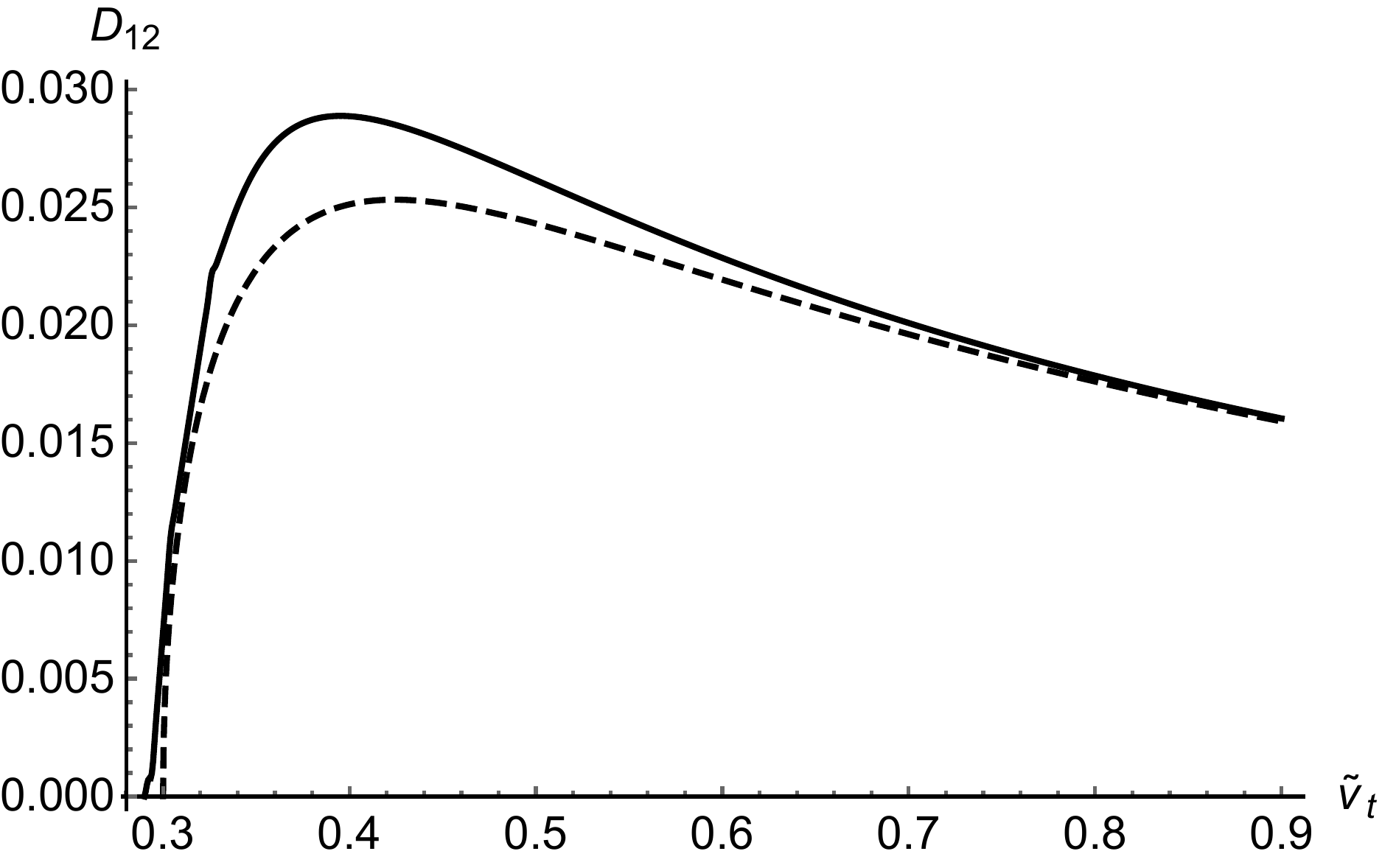}
\includegraphics[scale=0.3]{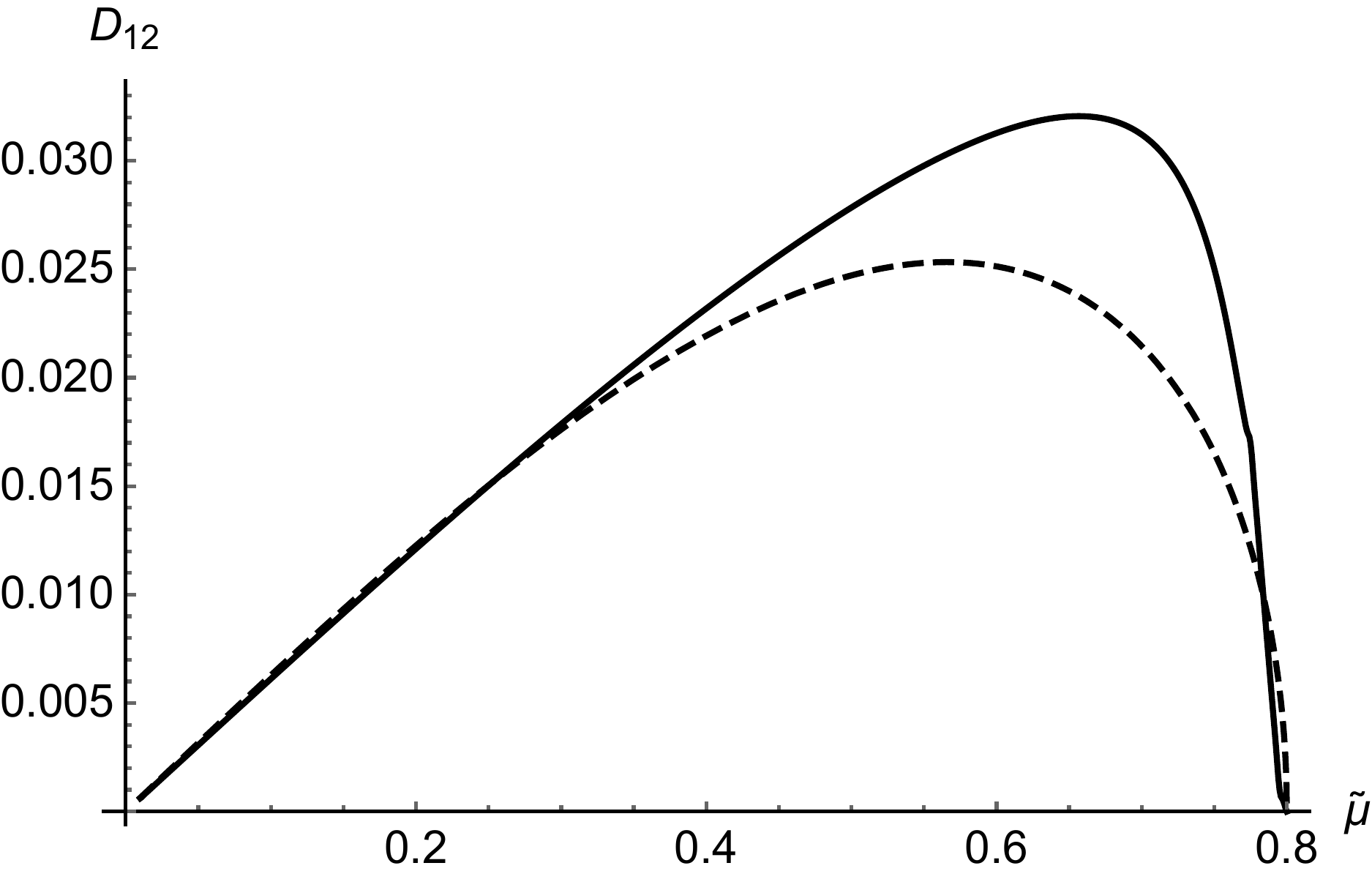}
\includegraphics[scale=0.3]{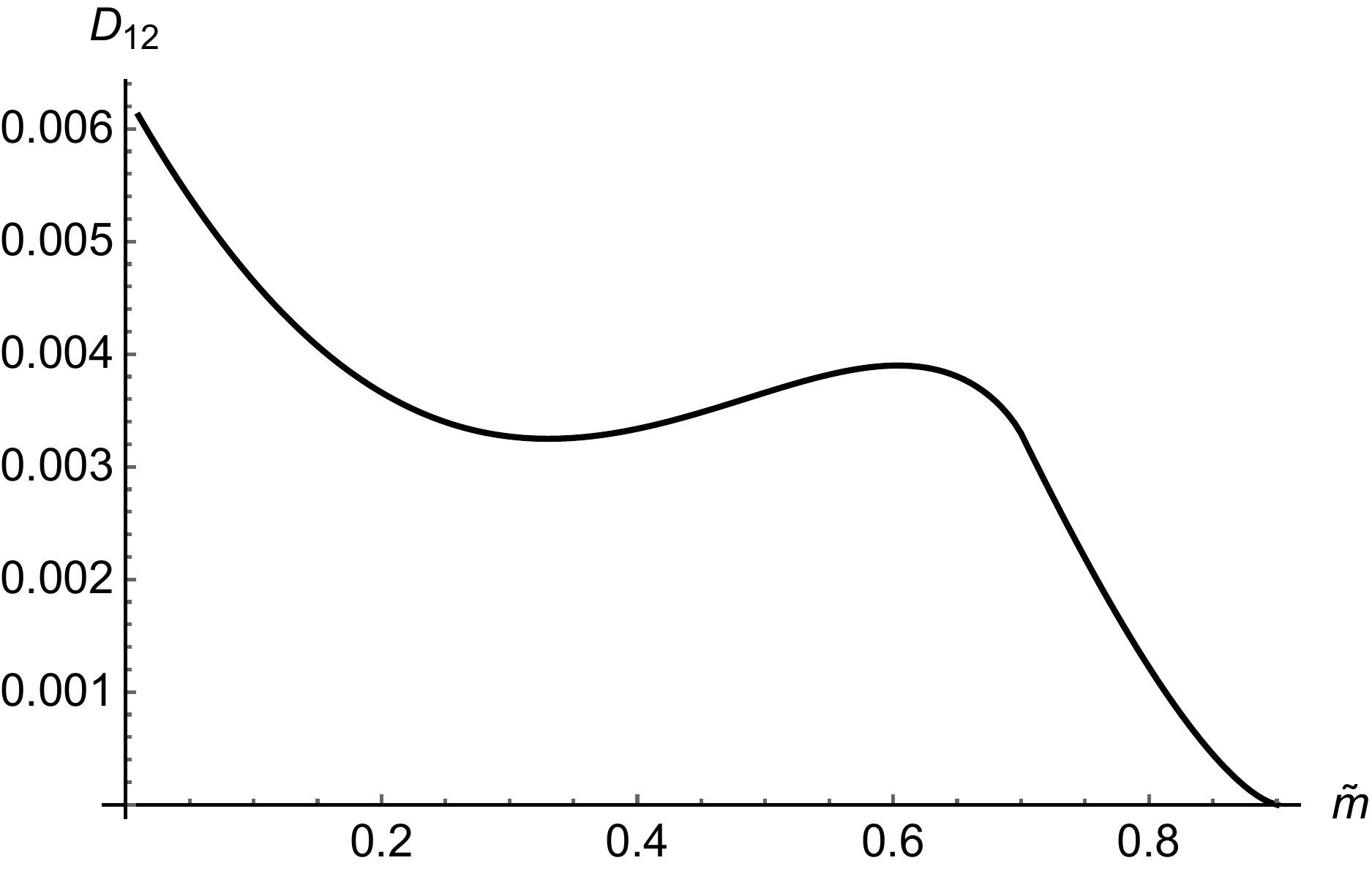}
\caption{Comparison of $D_{12}$ computed numerically (continuous lines) at $m\neq0$ and the analytical result (dashed lines), $m=0$. (a): $D_{12}$ as a function of $\tilde{v}_t$ for $\tilde{m}=0.01$ and $\tilde{\mu}=0.3$. (b): $D_{12}$ as a function of the dimensionless chemical potential $\tilde{\mu}$ for $\tilde{v}_t=0.8$ and $\tilde{m}=0.01$. (c): $D_{12}$ as a function of $\tilde{m}$ for the same values of $\tilde{\mu}$ and $\tilde{v}_t$ as before.} 
\label{fig:TiltedDnum}
\end{figure*}
We leave the full calculations of $D_{ab}$ to the appendix \ref{appendix:tiltD}. We quote here the final results. For $D_{12}$ we obtain:
\beq
D_{12}=\frac{\sign(m)}{2\pi^2}\left(\frac{\mu}{v_t k_0}\right)\sqrt{1-\frac{\mu^2}{v^2_t k^2_0}}\cos2\alpha.\label{D12}
\eeq
We stress here that we have calculated the \emph{total} $D_{12}$, integrated over the angle $\phi$. We see that the presence of a $\mathcal{P}$ breaking tilting term produces a non-zero result for the BCDM. In Fig. \ref{fig:TiltedDnum}(a,b) we compare the numerical results at finite $m$ with the analytical calculation with $m=0$. In Fig. \ref{fig:TiltedDnum}(c) we show the behavior of $D_{12}$ as a function of $m$. In the limit of zero mass, the numerical value matches the analytical result. As it happened with the linear Hall conductivity, at sufficiently large values of $m$ (for given values of $\mu>m$ and $v_t$), $D_{12}$ goes to zero. This can be understood along the same lines as in in Sec. \ref{sec:linear}: For some critical value of $m$, the conduction band covers the whole nodal loop, leading to a zero result due to integration over $\phi$.

We also obtain that $D_{21}=D_{12}$. In Appendix \ref{appendix:tiltD} we also provide details of the calculation of $D_{11}$($=-D_{22}$). This component takes the form
\beq
D_{11}= - \frac{\sign(m)}{2\pi^2}\left(\frac{\mu}{v_t k_0}\right)\sqrt{1-\frac{\mu^2}{v^2_t k^2_0}}\sin2\alpha.\label{D11}
\eeq

Collecting all the components of the BCDM, we can write it in a way similar to a quadrupole moment, in terms of the vectors $\hat{\bm{t}}$ and $\hat{\bm{v}}_t$:
\beqa
D_{ab}&=&\frac{\sign(m)}{4 \pi^2}\left(\frac{\mu}{v_t k_0}\right)\sqrt{1-\frac{\mu^2}{v^2_t k^2_0}}\cdot\nonumber\\
&\cdot&\left[ 2S_aS_b-S^2(\delta_{ab}-\delta_{a3}\delta_{b3}) \right] ,\label{quadrupD}
\eeqa
where the vector $\bm{S}$ is defined as $\bm{S} =  \hat{\bm{v}}_t+\hat{\bm{t}}\times\hat{\bm{v}}_t $. 

Inserting the BCDM (\ref{quadrupD}) into Eq. (\ref{def}) we find (again, in the collisionless regime)
\beqa
&&\boldsymbol{\mathcal{J}} ^{(0)} =\sign(m)\frac{e^3}{2 \pi ^{2} i \omega}\left(\frac{\mu}{v_t k_0}\right)\sqrt{1-\frac{\mu^2}{v^2_t k^2_0}}\cdot\nonumber\\
&\cdot&\left[\bm{\mathcal{E}}\times\bm{\mathcal{E}}^*- (\bm{S}\cdot\bm{\mathcal{E}})(\bm{S}\times\bm{\mathcal{E}}^*)-(\hat{\bm{t}}\cdot\bm{\mathcal{E}})(\hat{\bm{t}}\times\bm{\mathcal{E}}^*)\right],\label{TiltedHall2}
\eeqa
such that the real-valued non-linear Hall current is obtained as $\bm{J} ^{0} = \boldsymbol{\mathcal{J}} ^{(0)} +\boldsymbol{\mathcal{J}} ^{(0) \ast}$, where $\boldsymbol{\mathcal{J}} ^{(0) \ast}$ is the complex conjugate of (\ref{TiltedHall2}).

We can propose an effective electromagnetic action term that leads to the non-linear current $\bm{J} ^{0}$ in the same way as in the linear case for the current in Eq.(\ref{Halltilted}). In the low energy (long wavelength) limit the corresponding local effective action $\Gamma^{(2)}_H$ reads ($F_{\mu\nu}=\partial_{\mu}A_{\nu}-\partial_{\nu}A_{\mu}$ is the field strength):
\beq
\Gamma^{(2)}_H= - \int d^3\bm{x} dt \lambda^{H}_{\mu\nu\rho\alpha\beta} A_{\mu} \mbox{Im} \left( F_{\alpha\nu}F^{*}_{\beta\rho}  \right).
\eeq
In the low frequency regime, we might compare the term $\lambda^{H}_{\mu\nu\rho\alpha\beta}$ with the expression of the real-valued current $\boldsymbol{\mathcal{J}} ^{(0)} +\boldsymbol{\mathcal{J}} ^{(0) \ast}$, and obtain, after defining the vectors $S^{\mu}=(0, \bm{S})$, $\hat{t}^{\nu}=(0,\hat{\bm{t}})$, and using the Lorentz metric $\eta_{\rho\sigma}$:
\beqa
\lambda^H_{\mu\nu\rho\alpha\beta}&=& \frac{e^3}{\omega}\frac{\sign(m)}{2 \pi^2}\left(\frac{\mu}{v_t k_0}\right)\sqrt{1-\frac{\mu^2}{v^2_t k^2_0}}\eta_{0\alpha}\eta_{0\beta}\cdot\nonumber\\
&\cdot& \left(2\epsilon_{\mu\rho\sigma}S_{\nu}S^{\sigma}-S^2\epsilon_{\mu\nu\rho}+S^2\epsilon_{\mu\rho\sigma}\hat{t}^{\sigma}\hat{t}_{\nu}\right) .\label{coeffnonlinear}
\eeqa

This term is not topological in the same sense that it is the Chern-Simons term for the anomalous Hall conductivity, although it is directly related to the Berry curvature. Also, we have to remember that the BCDM $D_{ab}$ is a property of the Fermi surface and it is dissipative in origin (here we considered only the collisionless regime), in stark contrast to the linear Hall case that, although not acquiring quantized values, it is a non-dissipative property coming from both the filled bands and the Fermi surface\cite{H04,EMV09}. The remarkable result is that this non-linear response for tilted NLSM, although non topological, acquires contributions to the parity anomaly because it ultimately depends on the Berry curvature.
\section{Interband effects}
\label{sec:interband}
In previous sections we have studied the appearance of the parity anomaly in non-linear responses for frequencies $\omega$ smaller than $2\mu$, that is, neglecting interband transitions. However, it is known that the Berry curvature modifies the interband processes that can contribute to the non-linear optical responses\cite{BS80,SS00}. Moreover, it has been stated recently that it is possible to obtain a quantized circular photogalvanic response in shifted  Weyl semimetals\cite{JGM17}. In this section we study interband effects in the non-linear response of NLSM to see if they are sensible to the parity anomaly.
\subsection{Injection current}
\begin{figure*}
\includegraphics[scale=0.4]{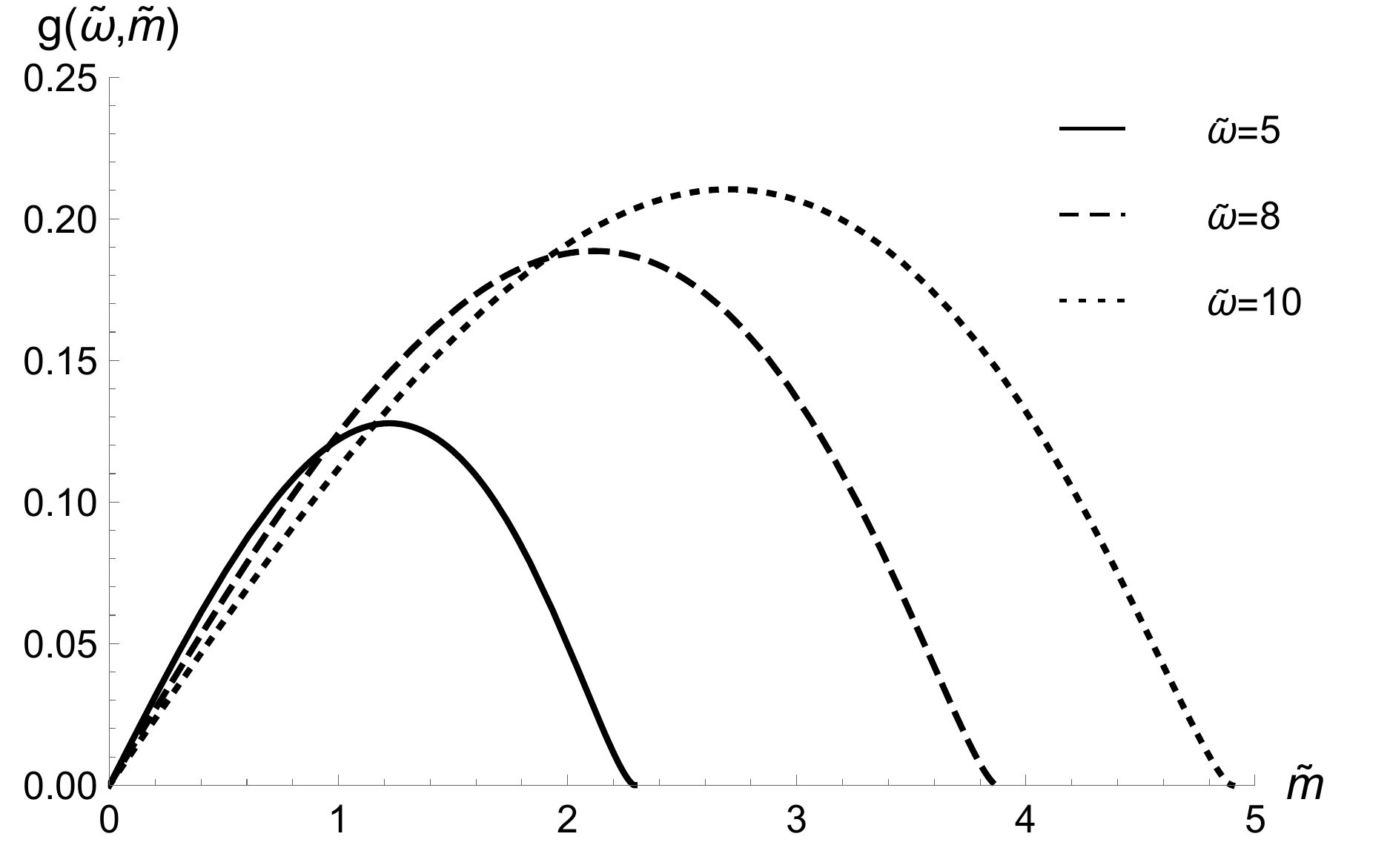}
\includegraphics[scale=0.4]{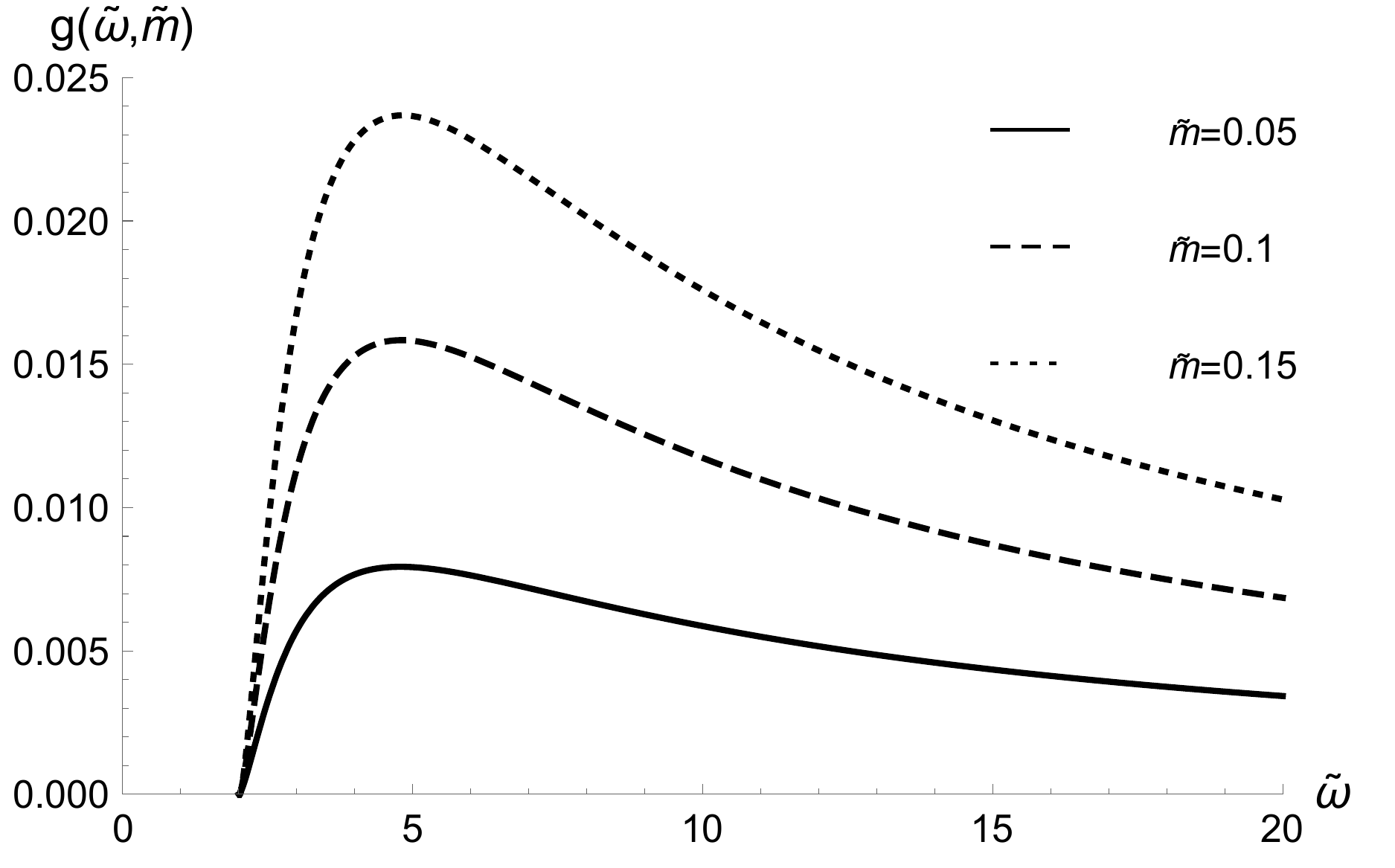}
\caption{(Left) Plot of $g (\tilde{\omega} , \tilde{m})$ as a function of $\tilde{m}$ for three different values of $\tilde{\omega}$. (Right) Plot of the function $g (\tilde{\omega} , \tilde{m})$ as a function of $\tilde{\omega}$ for three different values of $\tilde{m}$.} \label{fig:g-function}
\end{figure*}
The injection current is defined as the part of the photogalvanic effect that grows linearly with time:
\beq
\frac{d J_a}{dt}=\beta_{ab}(\omega)[\bm{\mathcal{E}}(\omega)\times\bm{\mathcal{E}}^*(\omega)]_b.\label{injectiondef}
\eeq
For a two-band model, as the one used in the present work, the tensor $\beta_{ab}(\omega)$ is defined as
\beq
\beta_{ab}(\omega)=\frac{i\pi e^3}{8\pi^3}\int d^3\bm{k}\frac{\partial E_{\bm{k},21}}{\partial k_a}\Omega_b\delta(\omega-E_{\bm{k},21}),\label{betadef}
\eeq
where $E_{\bm{k},21}$ stands for the difference between the conduction and valence dispersion relations, $E_{\bm{k},21}=\varepsilon_{2,\bm{k}}-\varepsilon_{1,\bm{k}}=2\varepsilon_{\bm{k}}$.

As in previous sections, in the pertinent quantities the polar angle $\phi$ factorizes in the integrals so we will write the tensor $\beta_{ab}$ as a function of this angle. Also, we will follow the strategy of computing $\beta^{\phi}_{ab}$ for non-zero values of $m$, and analyze the limit $m\to 0$.
The angle dependence can be casted in terms of the components of the vectors $\hat{\bm{r}}$ and $\hat{\bm{t}}$, so the tensor $\beta^{\phi}_{ab}(\omega)$ will take the following angular structure:
\beqa
\beta^{\phi}_{ab}(\omega)=ie^3g(\omega,m)\hat{r}_a\epsilon_{bcd}\hat{r}_c \hat{t}_d,\label{anglebeta}
\eeqa

It is possible to obtain an analytical expression of the function $g(\omega,m)$  (see the details in Appendix \ref{appendix:gfunction}). In terms of the dimensionless variables $\tilde{\omega}=\Lambda\omega/k^2_0$ and $\tilde{m}=\Lambda m/k^2_0$, this function reads:
\beqa
g(\tilde{\omega},\tilde{m})&=& -\frac{4\tilde{m}}{\tilde{\omega}^3} \Big\{ \sqrt{ (\tilde{\omega} / 2) ^{2} -\tilde{m}^2-1} - \left[ (\tilde{\omega} / 2) ^{2} -\tilde{m}^2  \right] \nonumber\\
& & \times \arctan \left( \sqrt{(\tilde{\omega} / 2) ^{2} -\tilde{m}^2-1} \right) \Big\} .\label{fullg}
\eeqa
In Fig. \ref{fig:g-function} we have plotted the function $g(\tilde{\omega},\tilde{m})$. In the left panel we observe the evolution of $g(\tilde{\omega},\tilde{m})$ for several values of $\tilde{\omega}$ varying $\tilde{m}$. The most relevant information we extract is that this function vanishes when the symmetry breaking parameter approaches to zero, even for a given angle $\phi$. This is in stark contrast to the case of the BCDM, where, for a given $\phi$, it remained non zero at $m=0$. This implies that the injection current \emph{does not} receive a finite contribution from the parity anomaly. In the right panel of Fig. \ref{fig:g-function} we have plotted $g(\tilde{\omega},\tilde{m})$ for several values of $\tilde{m}$, varying $\tilde{\omega}$. While the most salient feature of the function $g(\tilde{\omega},\tilde{m})$ is that it vanishes at $m=0$, it also vanishes for frequencies $\tilde{\omega}\geq 2\sqrt{\tilde{m}^2+1}$.

Although the injection current does not receive contributions from the parity anomaly, it is finite at non-zero symmetry breaking parameter $m$, even after integrating over the angle $\phi$. It is instructive to see the behavior of (\ref{injectiondef}) when circularly polarized light is applied to the NLSM. For right circular polarization, we have $\bm{\mathcal{E}}\times\bm{\mathcal{E}}^*=-i|\bm{\mathcal{E}}|^2 \hat{\bm{n}}$, where $\hat{\bm{n}}$ is the unit vector normal to the polarization plane. Inserting this in (\ref{anglebeta}) and integrating over $\phi$ we have, in vector form:
\beq
\frac{d\bm{J}}{dt}=\pi e^3g(\tilde{\omega},\tilde{m})|\bm{\mathcal{E}}|^2\hat{\bm{t}}\times\hat{\bm{n}},\label{injcurrent}
\eeq
that is, provided that $m$ is different from zero, an injection current appears perpendicular to the normal to the polarization plane \emph{and} perpendicular to the vector normal to the plane containing the nodal loop.

From an experimental perspective, the current in Eq.(\ref{injcurrent}) appears to grow indefinitely with time. This does not occur in real settings, and after some characteristic time $\tau$ the interband processes should equilibrate with other dissipative effects, leading to a steady current
\beq
\bm{J}_{\tau}=\tau\pi e^3g(\tilde{\omega},\tilde{m})|\bm{\mathcal{E}}|^2\hat{\bm{t}}\times\hat{\bm{n}}.\label{injcurrent2}
\eeq

\subsection{Shift current}
Another non-linear optical response that comes from inter-band transitions and it is susceptible of being modified by the Berry curvature is the shift current, defined through $J^{shift}_a=\sigma_{abb}(\omega)\mathcal{E}_b(\omega)\mathcal{E}_{b}(\omega)$. The tensor $\sigma_{abb}$ can be computed through the formula
\beq
\sigma_{abb}(\omega)={e^3}{2\pi^2}\int d^3\bm{k}\sum_{s\neq s\pr}f^{s s\pr}_0I^{abb}_{ss\pr}\delta(\omega-E_{s s\pr}),\label{shiftcond}
\eeq
where, as before, $E_{s s\pr}$ is the difference between the conduction and valence bands in a two-band model, $f^{s s\pr}_0=f^s_0-f^{s\pr}_0$ is the difference between the equilibrium distribution functions corresponding to the bands $s$ and $s\pr$, and $I^{abb}_{ss\pr}=\mbox{Im}(r^a_{s\pr s}r^b_{s s\pr;a})$ is constructed from the interband matrix elements of the position operator, related to the off diagonal elements of the non-abelian Berry connection as $r^a_{s s\pr}=i\braket{s|\partial_{k_a}s\pr}\equiv \mathcal{A}^a_{s s\pr}$. The semicolon denotes the covariant derivative: $r^b_{s s\pr;a}=\partial_a r^b_{s s\pr}-i(\mathcal{A}^a_{s s\pr}-\mathcal{A}^a_{s s})r^b_{s s\pr}$.

In absence of tilt, we can write the model (\ref{Ham1}) as $H=\sum_i\sigma_i d_i(\bm{k})$, the quantity $I^{abb}_{12}$ can be written as
\beq
I^{abb}_{12}=-\frac{m}{4\varepsilon_{\bm{k}}}\sum_{ijl}\epsilon_{ijl}d_l \partial_b d_{i} \left( \partial^2_{ab}d_{j}-\partial_a d_{j}\frac{\partial_b\varepsilon_{\bm{k}}}{\varepsilon_{\bm{k}}} \right).\label{Idef}
\eeq
Noticing that $d_2=m$ and $d_3=v_3 k_3$, after some algebra, we obtain
\beq
I^{abb}_{12}=\frac{m v_3}{4\varepsilon^4_{\bm{k}}}\partial_b\varepsilon_{\bm{k}}(\partial_a d_{1}\delta_{b1}-\partial_b d_{1}\delta_{a1}),
\eeq
with the shorthand notation $\partial_a=\partial_{k_a}$.

Now, using $d_1=\frac{1}{\Lambda}(k^2_0-k^2_{\rho})$, we can evaluate the derivatives and perform the momentum integration. As before, we will consider first the case with fixed $\phi$ and later discuss the integral over this angle. By symmetry considerations, it is easy to see that the non-vanishing components of $\sigma^{\phi}_{abb}(\omega)$ are
\begin{subequations}
\beq
\sigma^{\phi}_{311}(\omega)=\frac{2e^3}{\omega} \cos^2\phi g(\tilde{\omega},\tilde{m}),
\eeq
\beq
\sigma^{\phi}_{322}(\omega)=\frac{2e^3}{\omega} \sin^2\phi g(\tilde{\omega},\tilde{m}),
\eeq
\end{subequations}
with $g(\tilde{\omega},\tilde{m})$ defined in (\ref{fullg}).

As in the case of the injection current, the shift current does not survive to the limit $m\to 0$, implying that the parity anomaly does not leave its imprint in the shift current, either, even at fixed $\phi$.

If we integrate over $\phi$, for finite values of the parameter $m$, the total shift current $J^{shift}_a$ is
\beq
\bm{J}^{shift}=\pi^2\frac{e^3}{\omega}g(\tilde{\omega},\tilde{m})(\mathcal{E}^2_1+\mathcal{E}^2_2)\hat{\bm{t}},\label{shiftcurrent}
\eeq
that is, while the injection current (\ref{injcurrent}) circulates along the direction dictated by the intersection between the polarization plane and the plane containing the nodal loop, the shift current (\ref{shiftcurrent}) is perpendicular to the nodal loop. Also, we note that there are other contributions to the shift current that come from kinematic reasons, anisotropies in the velocities, or anisotropies in the distribution function $f_{\bm{k}}$. These contributions are not related to the Berry curvature, so they are not expected to be modified by the parity anomaly, so we will no consider here.

To conclude this section, we comment on the effects that the tilting term might cause in the injection and shift currents through the quantities (\ref{betadef}) and (\ref{Idef}). As it is clear from previous sections, the tilt term does not modify the Berry connection and therefore the Berry curvature, but modify the dispersion relations for electrons and holes in the same manner. Then since the dispersion relations enter in these expressions through their differences, we conclude that the only possible way to modify these responses is through the changes in the distribution functions. However, we have seen that the interband effects do not show terms coming from the parity anomaly, so they will vanish in the limit $m=0$ even if we introduce the changes induced by the tilting term in the distribution functions, so we will not analyze these effects here.

\section{Conclusions}
\label{sec:conclussions}
In the present work we have shown how the parity anomaly appears in the non-linear response in $\mathcal{P}*\mathcal{T}$ symmetric NLSM. As it happens for the linear Hall conductivity in these systems, we find that, if the nodal loop is not tilted,  the parity anomaly appears in these non-linear Hall responses for finite values of the polar angle characterizing the nodal loop, but vanishes after integrating over this polar angle. This occurs because each point at the nodal loop possesses a inversion symmetric partner belonging to the nodal loop as well, so both contributions cancel out, leading to a net zero response. We have found that tilting the nodal loop preserves the $\mathcal{P}*\mathcal{T}$ symmetry, but breaks this balance between inversion symmetric points on the nodal loop, giving rise to a finite linear and non-linear Hall responses. While the parity anomaly shows up in the non-linear Hall responses for the range of frequencies where interband transitions can be neglected, it does not appear in these non-linear responses coming from interband transitions, as the injection and shift currents.

Here we have considered NLSM in absence of SOC, where the $\mathcal{P}*\mathcal{T}$ symmetry topologically stabilizes the nodal loop. In presence of SOC other combinations of symmetries might stabilize the nodal loop. Then, a pertinent question is if some phenomenon similar to appearance of linear and non-linear Hall responses due to the parity anomaly will appear there. In this case, the minimal number of bands involved is the description of NLSM with SOC is four. However, the bandstructure will consist in two high-energy gapped bands, and two low energy bands forming the nodal loop, with an effective model for these two bands similar to (\ref{Ham1}), although the momentum dependence of $d_1(\bm{k})$ might be linear. In this case, all the results obtained in the present work are still valid, and what remains is to add any possible term coming from the gapped, high-energy sector that otherwise does not depend on the chemical potential. These symmetries are also compatible with the presence of tilting of the nodal loops\cite{HJL17,AWJ17}.
\section{Acknowledgements} 
We thank Laszlo Oroszlany for comments on the manuscript.
A. M. was supported by the CONACyT postdoctoral Grant No. 234774.
A.C. acknowledges financial support through the MINECO/AEI/FEDER, UE Grant No. FIS2015-73454-JIN. and the Comunidad de Madrid MAD2D-CM Program (S2013/MIT-3007). 
\appendix
\section{Details of the computation of the Hall current (\ref{Halltilted})}
\label{appendix:Hallcurrent}

For finite tilt oriented forming an angle $\alpha$ with the  $\hat{\bm{k}}_x$ axis, $\bm{v}_t=v_t(\cos\alpha,\sin\alpha,v^t_3)$, both valence and conduction bands contribute to the Hall conductivity in (\ref{Halldef}):

\beq
\sigma^{H}_{ab}=\sum_{s=\pm}\frac{e^2}{8\pi^3}\int d^3\bm{k}f^{s}_0(\bm{k})\epsilon_{abc}\Omega^s_{c}.\label{Halldef2}
\eeq
In the massless limit, we will use the expression (\ref{Berryzero}) for $\bm{\Omega}^s_{\bm{k}}$, and at zero temperature, the Dirac deltas in the expression for the Berry curvature simplify the distribution functions for electrons and holes in (\ref{Halldef2}) as
\begin{subequations}
\beq
f^{+}_0=\Theta(\mu-v_t k_0\cos(\phi-\alpha)),\label{distelectrons}
\eeq
\beq
f^{-}_0=1-\Theta(\mu-v_t k_0\cos(\phi-\alpha)).\label{distholes}
\eeq
\end{subequations}
The valence band (\ref{distholes}) contributes with two terms. The first one gives zero after integration over $\phi$. The second term goes with an extra minus sign with respect to (\ref{distelectrons}). This sign cancels out the extra minus sign from the Berry curvature $\bm{\Omega}^s_{\bm{k}}$ for the valence band, implying that in this case the contributions of electrons and holes add up, instead of cancelling each other:
\beq
\sigma^{H}_{ab}=\frac{e^2k_0}{4\pi^2}\sign(m)\epsilon_{abc}\int d\phi\Theta(\mu-v_t k_0\cos(\phi-\alpha))\hat{\bm{e}}^\phi_{c}.\label{Halldef3}
\eeq
From this expression it becomes clear that the integration domain of the polar variable $\phi$ is not the whole nodal loop but those values of $\phi$ that satisfy
$\cos(\phi-\alpha)\leq \mu/v_t k_0$, and the integration limits in (\ref{Halldef3}) are
\beq
\phi_{\pm}\equiv\alpha\pm\phi_0=\alpha\pm\arccos\left(\frac{\mu}{v_t k_0}\right).\label{intlimits}
\eeq
These angles are clearly depicted in Fig. \ref{cross}a for the case $\alpha = 0$ (tilting along the horizontal axis). If we choose $\bm{E}=E_3\hat{\bm{z}}$, it is easy to perform the integral in (\ref{Halldef3}) with the previous integration limits, obtaining
\begin{subequations}
\beq
\sigma^{H}_{13}=-\frac{e^2}{4\pi^2}k_0\sign(m)\sqrt{1-\frac{\mu^2}{v^2_t k^2_0}}\cos\alpha,
\eeq
\beq
\sigma^{H}_{23}=-\frac{e^2}{4\pi^2}k_0\sign(m)\sqrt{1-\frac{\mu^2}{v^2_t k^2_0}}\sin\alpha.
\eeq
\end{subequations}

If we denote $\hat{\bm{t}}=(0,0,1)$ as the normal vector to the plane containing the nodal loop in momentum space, we have $\bm{E}\times(\hat{\bm{v}}_t\times\hat{\bm{t}})=(-\cos\alpha E_3, -\sin\alpha E_3, 0)$ so we can compactly write
\beq
\bm{J}_{H}=\frac{e^2}{2\pi^2}\sign(m)k_0\sqrt{1-\frac{\mu^2}{v^2_t k^2_0}}\bm{E}\times(\hat{\bm{v}}_t\times\hat{\bm{t}}).\nonumber
\eeq
which is the expression (\ref{Halltilted}).
\section{Computation of the functions $X(m,\mu)$, $X_0(m,\mu)$, and $Q(m,\mu)$}. \label{appendix:XandQ}

The functions $X(m,\mu)$ and $Q(m,\mu)$ originate from the derivatives of the Berry curvature at finite $m$ ( the expression (\ref{BerryC})). $Q(m,\mu)$ comes from
\beqa
\frac{\partial\Omega_1}{\partial k_1}&=&- \frac{m v _{3}}{\Lambda} k _{2} \frac{\partial}{\partial k _{1}} \left( \frac{1}{\varepsilon_{\bm{k}} ^{3}} \right) =\nonumber\\
&-&\frac{6 m v _{3}}{\Lambda ^{3}} \frac{k^2 _\rho \sin\phi\cos\phi}{\varepsilon _{\bm{k}} ^{5}} \left( k _{0} ^{2} - k _{\rho} ^{2} \right) .
\eeqa
Separating the angular part, the integration of this expression over $k_\rho dk_\rho$ and $dk_3$ gives $Q(m,\mu)$.
On the other hand, 
\beqa
\frac{\partial\Omega_2}{\partial k_1}&=&\frac{m v_3}{\Lambda}\left(\frac{1}{\varepsilon^3_{\bm{k}}}-\frac{6k^2_\rho\cos^2\phi(k^2_0-k^2_\rho)}{\Lambda^2 \varepsilon^5_{\bm{k}}}\right).
\eeqa
The first term of the right hand side is the origin of the function $X(m,\mu)$, and again we see in the second term the function $Q(m,\mu)$.

The approximate function $X_0(m,\mu)$ comes from approximating the integral in $X(m,\mu)$ for values of $\mu$ slightly above the gap. Then we can approximate $\varepsilon_{\bm{k}}\simeq\sqrt{m^2+v^2_3 k^2_3+(2k_0 q_\rho/\Lambda)^2}$, with $q_\rho=k_\rho-k_0$:
\begin{align}
X _{0} (m , \mu) = \frac{m v _{3}}{8 \pi ^{3} \Lambda} \int f_0 \frac{k _{0} + q _{\rho}}{(m^2+v^2_3 k^2_3+(\frac{2k_0}{\Lambda}q_\rho))^{3/2}} d q _{\rho} d k _{3}.
\end{align}
The distribution function $f_0$ limits the integration to the range $m<\varepsilon_{\bm{k}}<\mu$. With the following change of variables, $q_\rho=\frac{\Lambda}{2k_0}r \cos\theta$ and $k_3=\frac{1}{v_3}r\sin\theta$,we have, after integrating over $\theta$:
\beqa
X_0(m,\mu)&=&\frac{m}{8\pi^2}\int^{\sqrt{\mu^2-m^2}}_0 \frac{r}{(m^2+r^2)^{3/2}}dr \nonumber\\
&=&\frac{1}{8\pi^2}\sign(m)\left(1-\frac{|m|}{\mu}\right),
\eeqa
which is the result (\ref{funcX0}).
\section{Computation of the components of the BCDM at finite tilting}
\label{appendix:tiltD}
We start with the detailed computation of $D_{12}$. In cylindrical coordinates, we have
\beq
D_{12}=\frac{1}{8\pi^3}\sum_{s=\pm}\int f^{s}_0(\varepsilon_{\bm{k}})\frac{\partial\Omega^{s}_{2}}{\partial k_1}k_\rho dk_{\rho}dk_3d\phi.\label{D12def}
\eeq
 Using the expression (\ref{Berryzero}) for $\bm{\Omega}_{\bm{k}} ^{s}$ and the expression of derivatives in cylindrical coordinates, $\partial_{k_1}=\cos\phi\partial_{k_{\rho}}-\sin\phi k^{-1}_{\rho}\partial_{\phi}$ we have
\beqa
D _{12} &=-&\frac{\sign(m)}{8\pi ^2}\sum_{s}s\int f^s_0(\varepsilon^s_{\bm{k}})\delta (k_3) ( k_{\rho}\cos ^2\phi\delta\pr(k _\rho-k_0)+\nonumber\\
&+&\sin^2\phi\delta(k_{\rho}-k_0) ) dk _{\rho}dk _3 d\phi,
\eeqa
where we have expressed $\delta\pr=\partial_{k_{\rho}}\delta$. To simplify further, we use the distributional definition of the derivative of the Dirac delta:
\beq
k_{\rho}f_0\delta\pr(k _\rho-k_0)=-\delta(k _\rho-k_0)(f_0+k_{\rho}f\pr_0),
\eeq
to obtain
\beqa
D_{12}&=&\sum_{s}s\frac{\sign(m)}{8\pi^2}\int dk_{\rho}dk_3 d\phi \delta(k_3)\delta(k_{\rho}-k_0)\cdot\nonumber\\
&\cdot&(f^s_0\cos2\phi-f^{s\prime}_0k_{\rho}\cos^2\phi)\nonumber\\
&\equiv& D^{(1)}_{12}+D^{(2)}_{12}.\label{D12}
\eeqa
The first piece $D^{(1)}_{12}$ is similar to the integral defining the linear Hall conductivity (see Appendix \ref{appendix:Hallcurrent}). We then use the expressions (\ref{distelectrons}) and (\ref{distholes}) for the conduction and valence band distribution functions that define the angular integration limits (\ref{intlimits}). As in the linear case, the opposite sign of the Berry curvature for electrons and holes cancels out with the opposite sign present in their distribution functions, so the contributions for electrons and holes add up leading, after integrating over $\phi$:
\beq
D^{(1)}_{12}=\frac{1}{2\pi^2}\sign(m)\left(\frac{\mu}{v_t k_0}\right)\sqrt{1-\frac{\mu^2}{v^2_t k^2_0}}\cos2\alpha.\label{D112}
\eeq
For the calculation of $D^{(2)}_{12}$, we need to compute the quantity $\sum_s s\partial_{k_{\rho}}f^s_0(\varepsilon^s(\bm{k}))=\sum_s s\partial_{k_{\rho}}\varepsilon^s(\bm{k})\partial_{\varepsilon}f^s_0(\varepsilon^s(\bm{k}))$, with
\beq
\frac{\partial \varepsilon^s(\bm{k})}{\partial k_{\rho}}=v_t \cos(\phi-\alpha)-s\frac{2k_{\rho}(k^2_0-k^2_\rho)}{\Lambda^2\sqrt{v^2_3 k^2_3+\frac{1}{\Lambda^2}(k^2_0-k^2_\rho)^2}},\label{intermediate}
\eeq
and $\partial_{\varepsilon}f_0(\varepsilon)=-\delta(\mu-\varepsilon)$. We insert the previous expressions for each $s$ in $D^{(2)}_{12}$, and, due to the extra minus sign from the Berry curvature in (\ref{Berryzero}), we obtain
\beqa
D^{(2)}_{12}&=&-\frac{\sign(m)k^2_0}{2\pi^2\Lambda}\int d\phi\int dk_\rho\sign(k_0-k_\rho)\delta(k_\rho-k_0)\cdot\nonumber\\
&\cdot&\delta(\mu-v_t k_0\cos(\phi-\alpha)).
\eeqa
We perform the integral over $k_\rho$ by substituting the Delta function by the Heat kernel regularization ($\mbox{Erf}$ is the error function):
\beqa
&&\int dk_\rho\sign(k_0-k_\rho)\delta(k_\rho-k_0)=\nonumber\\
&=&\lim_{a\to0}\frac{1}{a\sqrt{\pi}}\int dk_\rho\sign(k_0-k_\rho)e^{-\frac{(k_\rho-k_0)^2}{a^2}}=\nonumber\\
&=&\lim_{a\to0}\frac{1}{2}\left[\mbox{Erf}\left(\frac{k_0}{a}\right)-1\right]=0,
\eeqa
implying that $D^{(2)}_{12}=0$. This leaves us with the final result
\beq
D_{12}=\frac{\sign(m)}{2\pi^2}\left(\frac{\mu}{v_t k_0}\right)\sqrt{1-\frac{\mu^2}{v^2_t k^2_0}}\cos2\alpha.\label{D12}
\eeq

The $D_{21}$ component reads
\beq
D_{21}=\frac{1}{8\pi^3}\sum_{s=\pm}\int f^{s}_0(\varepsilon_{\bm{k}})\frac{\partial\Omega^{s}_{1}}{\partial k_2}k_\rho dk_{\rho}dk_3d\phi.\label{D21def}
\eeq
Following the same steps done in the main text, this component takes the intermediate form:
\beqa
D_{21}&=&-\sum_{s}s\frac{\sign(m)}{8\pi^2}\int dk_{\rho}dk_3 d\phi \delta(k_3)\delta(k_{\rho}-k_0)\cdot\nonumber\\
&\cdot&(f^s_0\cos2\phi+f^{s\prime}_0k_{\rho}\sin^2\phi).\label{D21}
\eeqa
The first term in the parenthesis is the same as $D^{(1)}_{12}$, while the second term is the same appearing in $D^{(2)}_{12}$ except for the replacement $\cos^2\phi\to\sin^2\phi$. $D^{(2)}_{12}$ vanishes independently of the angular dependence so does $D^{(2)}_{21}$, and we conclude that $D_{21}=D_{12}$.

Finally, let us evaluate the $D_{11}$ component:
\beqa
D_{11}&=&\frac{1}{8\pi^3}\sum_{s=\pm}\int f^s_0\frac{\partial\Omega^s_1}{\partial k_1}k_\rho dk_\rho dk_3d\phi=\nonumber\\
&=&-\frac{\sign(m)}{8\pi^2}\sum_{\pm}\int dk_\rho dk_3 d\phi f^s_0\delta(k_3)k_\rho\cdot\nonumber\\
&\cdot&\frac{\partial}{\partial k_1}[\sin\phi\delta(k_\rho-k_0)].
\label{D11def}
\eeqa
Applying the definition of $\partial_{k_1}$ in cylindrical coordinates and the definition of the derivative of the Delta function, we obtain
\beqa
D_{11}&=&\frac{\sign(m)}{8\pi^2}\sum_{s=\pm}\int dk_\rho dk_3d\phi \sin2\phi \delta(k_3)\cdot\nonumber\\
&\cdot&[2 f^s_0+k_\rho f^{s\prime}_0]\delta(k_\rho-k_0).
\eeqa
As before, the integral involving $f^{s\prime}_0$ vanishes, so
\beqa
D_{11}&=&\frac{\sign(m)}{4\pi^2}\sum_{s=\pm}\int d\phi \sin2\phi dk_3\delta(k_3)f^s_0\delta(k_\rho-k_0)dk_\rho=\nonumber\\
&=&-\frac{\sign(m)}{2\pi^2}\left(\frac{\mu}{v_t k_0}\right)\sqrt{1-\frac{\mu^2}{v^2_t k^2_0}}\sin2\alpha,\label{D11}
\eeqa
after integrating over $\phi$ with the integration limits (\ref{intlimits}). 
It is immediate to see that $D_{22}=-D_{11}$.
\section{Computation of the function $g(\omega,m)$}
\label{appendix:gfunction}
Let us evaluate the function $g(\omega , m )$ for an arbitrary frequency. We start with the definition 
\beqa
g(\omega , m) &=& -\frac{2 m v _{3}}{\Lambda ^{3}} \int  k _{\rho} d k _{\rho} d k _{3} \cdot\nonumber\\
&\cdot&\Theta (k _{\rho})\frac{k_\rho ^2 ( k _{0} ^{2} - k _{\rho} ^{2})}{\varepsilon _{\bm{k}} ^{4}} \delta \left(\omega - 2 \varepsilon _{\bm{k}} \right).
\eeqa
First, we express the Dirac delta function in terms of the roots of the argument. To this end, we have to employ the formula 
\beqa
\delta (f(x)) = \sum _{i} \frac{\delta (x - x _{i})}{\vert f ^{\prime} (x _{i}) \vert },
\eeqa
for an arbitrary (continuously differentiable) function $f(x)$ with its roots $x _{i}$ giving $f (x _{i}) = 0$. In the present case we have $\omega = 2 \varepsilon _{\bm{k}}$, wherefrom we obtain the roots
\beq
k _{\rho \pm\pr \pm} ^{\ast} = \pm\pr \sqrt{k _{0} ^{2} \mp \Lambda \sqrt{(\omega / 2) ^{2} - m ^{2} - v _{3} ^{2} k _{3} ^{2}} } .
\eeq
Since our integral is only for positive values of $k _{\rho}$ (it is a radius in polar coordinates), then we are left with only two roots, which corresponds to $k _{\rho \pm} ^{\ast} = k _{\rho + ^{\prime} \pm} ^{\ast} > 0$. Therefore we can write
\beq
\delta (\omega - 2 \varepsilon _{\bm{k}}) = \frac{\delta (k _{\rho} -  k _{\rho +} ^{\ast} )}{2 \vert \varepsilon _{\bm{k}}\pr \vert _{k _{\rho} = k _{\rho +} ^{\ast}} } \Theta (k _{\rho +} ^{\ast}) + \frac{ \delta ( k _{\rho} -  k _{\rho -} ^{\ast} )}{2 \vert \varepsilon _{\bm{k}} ^{\prime} \vert _{k _{\rho} = k _{\rho -} ^{\ast}}} \Theta (k _{\rho -} ^{\ast}).
\eeq
The constraint imposed by the step function defines the limits of integration over the variable $k _{3}$. We can also directly verify that
\beq
\frac{\partial \varepsilon _{\bm{k}}}{\partial k _{\rho}} \Bigg| _{k _{\rho} = k _{\rho \pm} ^{\ast}} = - \frac{2 k _{\rho \pm} ^{\ast} \left( k _{0} ^{2} - k _{\rho \pm} ^{\ast \, 2} \right)}{\Lambda ^{2} \varepsilon_{\bm{k}} \vert _{k _{\rho \pm} ^{\ast}}} = - \frac{2 k _{\rho \pm} ^{\ast} \left( k _{0} ^{2} - k _{\rho \pm} ^{\ast \, 2} \right)}{\Lambda ^{2} (\omega / 2)} 
\eeq
and $\quad \varepsilon _{\bm{k}} \vert _{k _{\rho} = k _{\rho \pm} ^{\ast}} =\omega / 2$.

Inserting these results into the integral expression and integrating with respect to $k _{\rho}$:
\begin{widetext}
\beqa
g(\omega , m) &=& - \frac{2 m v _{3}}{\Lambda ^{3}}  \int \frac{k _{\rho} ^{3}}{\varepsilon _{\bm{k}} ^{4}} \left( k _{0} ^{2} - k _{\rho} ^{2} \right) \left[ \frac{\delta (k _{\rho} -  k _{\rho +} ^{\ast} )}{2 \vert \varepsilon_{\bm{k}}\pr \vert _{k _{\rho} = k _{\rho +} ^{\ast}} } \Theta (k _{\rho +} ^{\ast}) + \frac{ \delta ( k _{\rho} -  k _{\rho -} ^{\ast} )}{2 \vert \varepsilon _{\bm{k}}\pr\vert _{k _{\rho} = k _{\rho -} ^{\ast}}} \Theta (k _{\rho -} ^{\ast}) \right] d k _{\rho} d k _{3}=\nonumber\\
 &=& - \frac{4 m v _{3}}{\Lambda} \frac{1}{ \omega^{3}} \int \left[ k _{\rho +} ^{\ast \, 2} \Theta (k _{\rho +} ^{\ast}) - k _{\rho -} ^{\ast \, 2} \Theta (k _{\rho -} ^{\ast}) \right] \, d k _{3} .
\eeqa
Now we determine the integration region for $k _{3}$. On the one hand, the condition $k _{\rho +} ^{\ast} = 0$ yields $k _{3} \in [- k _{3} ^{\ast} , + k _{3} ^{\ast}]$, where $k _{3} ^{\ast}=\frac{1}{v _{3}} \sqrt{(\omega / 2) ^{2} - m ^{2} - (k _{0} ^{2} / \Lambda ) ^{2}}$.

On the other hand, we can see that the condition $k _{\rho -} ^{\ast} = 0$ has no real solutions for $k _{3}$. Therefore, we are only left with the first integral:
\beqa
g (\omega , m) =
- \frac{4 m v _{3}}{\Lambda\omega^{3}}\int _{- k _{3} ^{\ast}} ^{+ k _{3} ^{\ast}} \left[ k _{0} ^{2} - \Lambda \sqrt{(\omega / 2) ^{2} - m ^{2} - v _{3} ^{2} k _{3} ^{2}} \right]  d k _{3} .
\eeqa
This integral can be performed in a simple fashion. The result is
\beqa
&&g(\omega , m) =\nonumber\\
&=& -\frac{4 m v _{3}}{\Lambda \omega^3}\left\lbrace 2 k _{0} ^{2} k _{3} ^{\ast} - \frac{\Lambda}{v _{3}} \left[ v _{3} k _{3} ^{\ast} \sqrt{(\omega / 2) ^{2} - m ^{2} - v _{3} ^{2} k _{3} ^{\ast 2}} +[ (\omega / 2) ^{2} - m ^{2}] \arctan \left( \frac{v _{3} k _{3} ^{\ast}}{\sqrt{(\omega / 2) ^{2} - m ^{2} - v _{3} ^{2} k _{3} ^{\ast \, 2}}} \right) \right] \right\rbrace . 
\eeqa

After some algebraic simplifications and defining the dimensionless quantities $\tilde{m}$ and $\tilde{\omega}$, we obtain
\beqa
g ( \tilde{\omega} , \tilde{m} ) =
 -\frac{4\tilde{m}}{\tilde{\omega}^3} \Big\{ \sqrt{ (\tilde{\omega} / 2) ^{2} -\tilde{m}^2-1} - \left[ (\tilde{\omega} / 2) ^{2} -\tilde{m}^2  \right] \arctan \left( \sqrt{(\tilde{\omega} / 2) ^{2} -\tilde{m}^2-1} \right) \Big\}
\eeqa
\end{widetext}

\end{document}